\documentclass[11pt]{article}
\usepackage{jheppub}
\usepackage{amsmath,amssymb,amsfonts,graphicx,slashed,color}

\newcommand{\myfig}[3]{
	\begin{figure}[ht]
	\centering
	\includegraphics[width=#2cm]{#1}\caption{#3}\label{fig:#1}
	\end{figure}
	}
\newcommand{\myfigtab}[4]{
	\begin{figure}[ht]
	\centering
	\begin{tabular} {c c c}
	\includegraphics[width=#3cm]{#1} & \hspace{0.8cm} & \includegraphics[width=#3cm]{#2}
	\end{tabular}
	\caption{#4}\label{fig:#1}
	\end{figure}
	}

\renewcommand{\a}{\alpha}
\renewcommand{\b}{\beta}
\renewcommand{\c}{\gamma}
\renewcommand{\d}{\delta}

\newcommand{\pa}{\partial}

\newcommand{\tr}{{\rm tr}}
\newcommand{\be}{\begin{equation}}
\newcommand{\ee}{\end{equation}}
\newcommand{\bea}{\begin{eqnarray}}
\newcommand{\eea}{\end{eqnarray}}
\newcommand{\beq}{\begin{equation}}
\newcommand{\eeq}{\end{equation}}
\newcommand{\beqn}{\begin{eqnarray}}
\newcommand{\eeqn}{\end{eqnarray}}

\newcommand{\bs}{\boldsymbol}

\newcommand{\bdx}{\mathbf{x}}

\newcommand{\cO}{\mathcal{O}}

\newcommand{\he}{\epsilon}
\newcommand{\vx}{\mathbf{x}}
\newcommand{\cW}{\mathcal{V}}
\newcommand{\cL}{\mathcal{L}}

\newcommand{\na}{\nabla}

\usepackage{upgreek}
\newcommand{\pert}{\upvarepsilon}

\title{Nonlinear Gravity from Entanglement \\ in Conformal Field Theories}

\author[a]{Thomas Faulkner}
\author[b]{\!, Felix M. Haehl}
\author[b]{\!, Eliot Hijano}
\author[c]{\!, Onkar Parrikar}
\author[c,d]{\!, Charles Rabideau}
\author[b]{\! Mark Van Raamsdonk}

\affiliation[\,a]{Department of Physics, University of Illinois,\\
 1110 W. Green St., Urbana IL 61801-3080, U.S.A.}
\affiliation[\,b]{Department of Physics and Astronomy, University of British Columbia,\\
6224 Agricultural Road, Vancouver, B.C.\ V6T 1Z1, Canada.}
\affiliation[\,c]{David Rittenhouse Laboratory, University of Pennsylvania,\\
209 S.33rd Street, Philadelphia PA, 19104, U.S.A.}
\affiliation[\,d]{Theoretische Natuurkunde, Vrije Universiteit Brussel (VUB), and \\ International Solvay Institutes, Pleinlaan 2, B-1050 Brussels, Belgium}

\emailAdd{tomf@illinois.edu}
\emailAdd{f.m.haehl@gmail.com}
\emailAdd{ehijano@phas.ubc.ca}
\emailAdd{parrikar@sas.upenn.edu}
\emailAdd{rabideau@sas.upenn.edu}
\emailAdd{mav@phas.ubc.ca}

\abstract{
In this paper, we demonstrate the emergence of nonlinear gravitational equations directly from the physics of a broad class of conformal field theories. We consider CFT excited states defined by adding sources for scalar primary or stress tensor operators to the Euclidean path integral defining the vacuum state. For these states, we show that up to second order in the sources, the entanglement entropy for all ball-shaped regions can always be represented geometrically (via the Ryu-Takayanagi formula) by an asymptotically AdS geometry. We show that such a geometry necessarily satisfies Einstein's equations perturbatively up to second order, with a stress energy tensor arising from matter fields associated with the sourced primary operators. We make no assumptions about AdS/CFT duality, so our work serves as both a consistency check for the AdS/CFT correspondence and a direct demonstration that spacetime and gravitational physics can emerge from the description of entanglement in conformal field theories.
}

\keywords{}

\begin{document}

\maketitle

\parskip=10pt

\section{Introduction}

The AdS/CFT correspondence suggests that certain quantum field theories can be reinterpreted to provide a fundamental definition of some corresponding quantum theories of gravity. For these field theories, each quantum state has a dual gravitational interpretation; it encodes some asymptotically AdS spacetime whose boundary geometry is equivalent to the fixed spacetime on which the field theory lives. These spacetimes should satisfy the dynamical constraints of the corresponding gravitational theory; at the classical level these are the Einstein equations or some other covariant gravitational equations.

A few very basic questions about the correspondence are:
\begin{enumerate}
\item
For which field theories does a dual gravitational interpretation exist?
\item
Which states of these theories have a dual gravitational interpretation well-described by a classical spacetime?
\item
How is the spacetime geometry encoded in the field theory state in this case?
\item
What CFT physics implies that these encoded geometries satisfy the expected dynamical constraints?
\end{enumerate}
Despite a great deal of evidence for the validity of the correspondence, our understanding of these questions is far from complete. However, significant progress has come recently from the realization \cite{Maldacena:2001kr, Ryu:2006bv, Swingle:2009bg, VanRaamsdonk:2009ar, VanRaamsdonk:2010pw} that the spacetime geometry and gravitational physics encoded by the field theory is intimately connected with the structure and dynamics of quantum entanglement in the field theory state.

Central to the recent developments is the proposal by Ryu and Takayanagi \cite{Ryu:2006bv, Ryu:2006ef} (and its covariant generalization by Hubeny, Rangamani and Takayanagi \cite{Hubeny:2007xt} - we will refer to this as the HRRT formula) that the entropy of a spatial subsystem of the field theory, which measures the entanglement of that subsystem with the rest of the system, has a direct geometrical interpretation as the area of a certain surface in the dual spacetime. Through this connection, we can in principle deduce much of the spacetime geometry associated with a state by seeking a geometry that correctly reproduces the CFT entanglement entropies.

Starting from the HRRT proposal, it has also been possible to begin to understand the field theory origin of the dynamical constraints on the dual spacetimes. In \cite{Lashkari:2013koa, Faulkner:2013ica}, it was demonstrated that any spacetime geometry that correctly captures the entanglement entropy of a near-vacuum CFT state to first order in the perturbation must satisfy Einstein's equations linearized about Anti de Sitter space.\footnote{  A different derivation of the same result was given more recently in \cite{Mosk:2016elb,Czech:2016tqr} using the technology of kinematic space and OPE blocks. The derivation was extended to include bulk quantum corrections in \cite{Swingle:2014uza}.}

The main goal of this paper is to extend these results to the nonlinear level. Here, one has to be careful to work with states that are expected to have a good classical description on the gravity side, as opposed to states that might describe quantum superpositions of different geometries. After motivating a class of CFT states with this property, we will show that with only a minimal assumption about the CFT, a spacetime which captures via the HRRT formula the entanglement entropy of such a state up to second order in perturbations around the vacuum state must satisfy Einstein's equations perturbed to second order about AdS. The stress tensor acting as a source is that for matter fields determined by the non-zero one-point functions in the CFT state. Our calculations make use of techniques developed in  \cite{Faulkner:2014jva} (see also \cite{Faulkner:2015csl, Faulkner:2016mzt}) for perturbative CFT calculations of entanglement entropy and methods developed in \cite{Hollands:2012sf, Lashkari:2015hha} for the relevant gravity calculations.

In demonstrating the necessity of Einstein's equations, we need only demand that the spacetime in question captures the entanglement entropies for ball-shaped regions (of various sizes and in various frames of reference) of the CFT. A surprising by-product of our analysis is that a spacetime satisfying Einstein's equations and capturing the entanglement entropy for ball-shaped regions always exists as long as the CFT satisfies a single constraint relating the overall coefficient $C_T$ in the stress-tensor two-point function to an overall coefficient $a^*$ appearing in the vacuum entanglement entropy for a ball. This is a manifestation of the fact that to second order in the state deformation, the ball-entanglement entropies expressed in terms of CFT one-point functions are given by a nearly universal formula, depending only on the coefficients $C_T$ and $a^*$. Thus, if the ball entanglement entropies at this order are geometrical for certain CFTs, they must be geometrical for all CFTs with the same relation between these coefficients.

In fact, we will suggest that for {\it any} CFT, it is always possible to represent the ball-entanglement entropies geometrically if we replace the area functional in the HRRT formula with some particular functional in a one parameter family. In this case, the geometry calculating entanglement entropies with this functional must satisfy the equations of motion associated with some one-parameter family of gravitational actions.

We have stated above a bare-bones version of the results of this paper; before getting into the technical calculation, we will now give a somewhat more detailed summary of our results and a brief summary of their derivation.

\subsection{Summary of results}

We consider general conformal field theories on $d$ dimensional Minkowski spacetime. For any such CFT, we can define two parameters $a^*$ and $C_T$, both of which will be referred to as central charges since in even dimensions they can be related to the $a$- and $c$-central charges parameterizing the trace anomaly. The parameter $a^*$ governs the overall normalization of the universal parts of the vacuum entanglement entropy for ball-shaped regions \cite{Ryu:2006ef,Solodukhin:2008dh,Myers:2010xs, Casini:2011kv}; apart from this overall parameter, the ball entanglement entropies for all theories in a given dimension are identical. The second parameter $C_T$ gives the overall normalization of the CFT stress tensor two-point function \cite{Osborn:1993cr,Erdmenger:1996yc}. Again, these two-point functions take a universal form up to the overall normalization. For the most part of this paper we will focus on the case where\footnote{ In even dimensions $a^*$ is precisely the coefficient $A$ of the Euler density in the trace anomaly for $\langle T^\mu_\mu \rangle$. The relation between the usual central charge $c$ and the stress-tensor two-point function normalization is $C_T = (40/\pi^4)c$ for $d=4$ (see, for example \cite{Myers:2010xs}). Then the condition \eqref{eq:ca} reduces to $c=a$.}
\beq \label{eq:ca}
 \widetilde{C}_T \equiv   \frac{\pi^d\,(d-1)}{\Gamma(d+2)} \, C_T =   a^* \,,
\eeq
where the first equality defines $\widetilde{C}_T$ in terms of the stress tensor two-point normalization $C_T$ (using the convention of \cite{Hung:2011nu}).
In the discussion section we return to cases where $\widetilde{C}_T = a^*$ does not hold.

Our results apply to CFT states of the form
\beq
\label{PIstate0}
\langle \varphi_{(0)}(\bdx)| \psi_{\lambda} (\pert) \rangle = \int^{\varphi(x^0_E=0,\bdx) = \varphi_{(0)}(\bdx)}[D \varphi] e^{- \int_{-\infty}^0 d x^0_Ed^{d-1}\bdx\,\left(\cL_{CFT}+\lambda_\alpha( x; \pert) {\cal O}_\alpha(x)\right)} \; .
\eeq
created by Euclidean path-integrals of the type discussed in \cite{Botta-Cantcheff:2015sav,Christodoulou:2016nej} and inspired by \cite{Skenderis:2008dh,Skenderis:2008dg} (see section \ref{sec:EuclPI} for further details). Here, $\lambda_\alpha(x ; \pert)$ are sources for various primary operators in the theory, which will be taken to vanish for the Euclidean time $x^0_E \to 0$. In the holographic context these operators will be the low-dimension operators dual to \emph{light} fields in the bulk, but our results apply to general CFTs. We will motivate this class of states by arguing that in the holographic case, there is a specific prescription to associate classical bulk Lorentzian spacetimes to states defined in this way. Roughly, we can think of these states as giving rise to coherent states of the perturbative quantum fields in AdS \cite{Botta-Cantcheff:2015sav}. In a separate paper \cite{MPRV}, evidence will be presented that any near-AdS spacetime (defined perturbatively about AdS) can be described in this way.

For these states, we will assume that the sources turn off for $\pert \to 0$ and consider the perturbative expansion of various field theory quantities in powers of $\pert$. Specifically, we define
\beq
S(x, R, u; \pert) = S^{(0)}(x, R, u) + \pert\, \delta S^{(1)} (x, R, u) + \pert^2 \delta S^{(2)} (x, R, u) + \dots
\eeq
to be the entanglement entropy of a ball-shaped region of radius $R$ centered at spacetime point $x^\mu$ in a spatial plane perpendicular to timelike vector $u^\mu$; this depends implicitly on the sources $\lambda_\alpha$.

Finally, we define
\noindent
{\bf The Hubeny-Rangamani-Ryu-Takayanagi (HRRT) formula:} {\it Given an asymptotically $AdS$ geometry $M$ with dimensionless metric,\footnote{ In the context of $AdS/CFT$ or a gravitational theory, the dimensionless metric is defined as the metric expressed in units of $G_N$.} the HRRT formula associates to any region $A$ on the boundary of $AdS$ the quantity
\beq
S^{grav}_{M}(A) = {1 \over 4}{\rm Area} (\widetilde{A}) \; ,
\eeq
where $\widetilde{A}$ is the minimal-area codimension-two spacelike surface in $M$ that is homologous to $A$ (which also implies that $A$ and $\widetilde{A}$ have the same boundary) and extremizes the area functional.}

In this paper, we will demonstrate the following:
\vskip 0.1 in
\noindent
{\bf Main results}

\begin{enumerate}
\item
{\it Consider any CFT with $\widetilde{C}_T = a^*$ in a state of the form (\ref{PIstate0}). Then there exists a geometry $M(\pert)$ defined perturbatively by a metric $g^{(0)}_{AdS} + \pert \,\delta g^{(1)} + \pert^2 \delta g^{(2)} + \dots$ that correctly computes entanglement entropies up to order $\pert^2$ for all ball-shaped regions via the HRRT formula.}
\item
{\it This geometry $M(\pert)$ satisfies Einstein's equations, up to second order in $\pert$, where the stress-energy tensor is that for a set of matter fields corresponding to the sourced operators ${\cal O}_\alpha$, and these matter fields solve linearized equations about AdS with boundary conditions specified by the CFT one-point functions of ${\cal O}_\alpha$.}
\end{enumerate}

These statements are motivated by, but do not assume any aspects of the AdS/CFT correspondence. Thus, they might be seen as a direct demonstration that gravitational physics (including bulk locality) emerges naturally in the description of entanglement in conformal field theories. That our results hold with such minimal assumptions on the CFT may be surprising, but we will see that this is a necessary byproduct of the existence of conformal field theories with Einstein gravity duals together with the universal form of two point functions in CFTs.

The geometry $M(\pert)$ should not in general be thought of as a genuine holographic dual to the CFT state; unless the CFT satisfies certain extra consistency conditions, $M(\pert)$ will fail to correctly compute other CFT quantities if we try to calculate them using the standard AdS/CFT dictionary.

As an example, for a CFT dual to some theory of gravity with higher curvature corrections, we do not expect that the spacetime $M(\pert)$ defined above and satisfying Einstein's equations is the correct dual. But our result shows that it captures the ball entanglement entropies to order $\pert^2$ via the ordinary HRRT formula. What must be true in this case is that there is another spacetime $M_{\cal L}(\pert)$ satisfying the equations of motion for an appropriate higher derivative Lagrangian such that $M_{\cal L}(\pert)$ correctly encodes the entanglement entropies via the entanglement entropy functional associated with ${\cal L}$. Thus, we expect that our results above can be generalized, with area replaced by some generalized entanglement functional associated with ${\cal L}$ and Einstein's equations replaced by the equations of motion associated with ${\cal L}$. We will discuss this more in the final section.

\subsection{Summary of the derivation}

We now provide a brief summary of our derivation. We begin by recalling that for the vacuum state of any CFT, the density matrix $\rho_A^{(0)}$ for a ball-shaped region $A$ is given explicitly by
\be
\label{modHam}
\rho_A^{(0)} = {1 \over Z} e^{- H_A} \qquad \qquad H_A \equiv \int_A d^d x \, \zeta^0(x) T_{00}(x)
\ee
where for a ball of radius $R$ and radial coordinate $r$, $\zeta^0(x) = \pi (R^2 - r^2)/R$. In other words, the density matrix is thermal with respect to $H_A$, known as the modular Hamiltonian for the region. This follows via a conformal transformation from the fact that the vacuum density matrix for a half-space in any Lorentz-invariant quantum field theory is thermal with respect to a boost generator \cite{Casini:2011kv}.

As shown by Casini, Huerta, and Myers \cite{Casini:2011kv}, the result (\ref{modHam}) implies that the vacuum entanglement entropy
\be
S_A^{(0)} = - \tr(\rho_A^{(0)} \log \rho_A^{(0)})
\ee
for a ball-shaped region in any CFT is the same for all CFTs up to an overall constant $a^*$. In even dimensions, the universal coefficient $a^*$ of the logarithmic term in the entanglement entropy agrees with the $A$-type trace anomaly coefficient \cite{Ryu:2006ef,Solodukhin:2008dh}. In odd dimensions, the universal term in the entanglement entropy is constant and its coefficient $a^*$ characterizes the sphere partition function of the CFT, viz., $a^* \propto \log Z_{S^d}$.
In any dimension, the universal ball entanglement entropy exactly agrees with the HRRT formula applied to pure AdS
\be
\label{AdS}
g^{(0)} = {\ell^2 \over z^2}(dz^2 + dx^\mu dx_\mu)
\ee
if we set the AdS radius as
\beq  \label{eq:astar}
 \ell = \left( \frac{\Gamma(d/2)}{\pi^{d/2}} \, 8\pi G_N \, a^* \right)^{\frac{1}{d-1}} \,.
\eeq

\subsubsection*{Key identities}

In order to establish that perturbed states also have ball-entanglement represented geometrically, and that the relevant geometries satisfy Einstein's equations, we make use of two key identities.

The first follows from the definition of {\it relative entropy} in the CFT. For a general state of the CFT, the relative entropy is a measure of how different the density matrix $\rho_A$ for region $A$ is from a reference density matrix $\rho^{(0)}_A$, which we will take to be the vacuum density matrix. It is defined as:
\be
S(\rho_A || \rho^{(0)}_A) = \tr(\rho_A \log(\rho_A)) - \tr(\rho_A \log(\rho^{(0)}_A)) \; .
\ee
From this definition, we have immediately that
\be
\label{RE}
\Delta S_A - \Delta \langle H_A\rangle= - S(\rho_A || \rho^{(0)}_A)
\ee
where $\Delta$ refers to the difference from quantities defined in the vacuum state.

\myfig{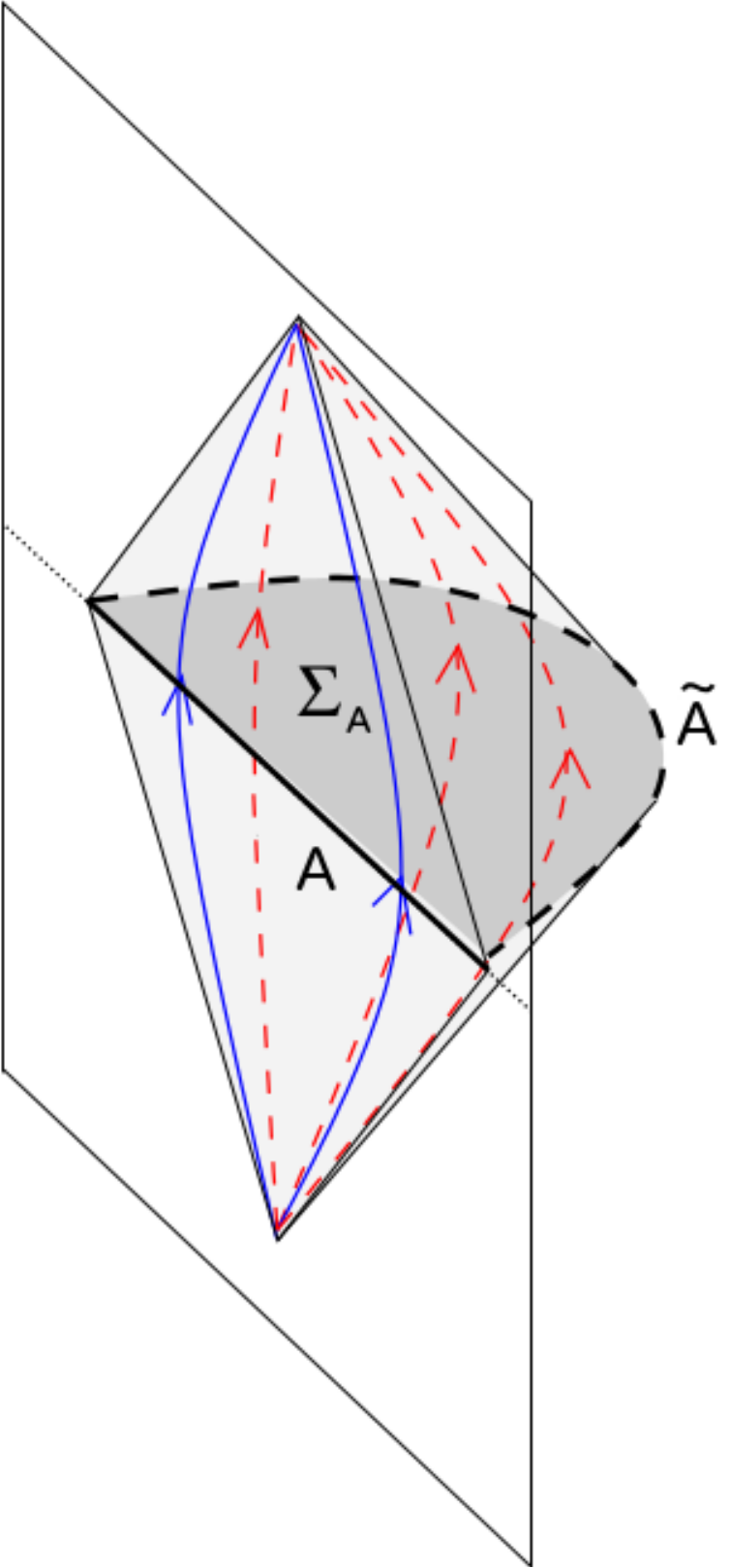}{4}{\small{\textsf{AdS-Rindler patch associated with a ball $A$ on a spatial slice of the boundary. Solid blue paths indicate the boundary flow associated with $H_A$ and the conformal Killing vector $\zeta_A$. Dashed red paths indicate the action of the Killing vector $\xi_A$.}}}

The second key identity is a geometrical identity that we will apply to one-parameter families of asymptotically AdS spacetimes defined by some metric $g(\pert)$ with matter fields $\phi_\alpha(\pert)$. To describe this, we define $\widetilde{A}$ to be the HRRT surface in pure AdS associated with a ball $A$ and $\Sigma_A$ to be the spatial region between $A$ and $\widetilde{A}$, as shown in figure \ref{fig:hyperbolic2.pdf}. The domain of dependence of the region $\Sigma_A$ forms a Rindler wedge of AdS. In this region, there is a Killing vector $\xi_A$ which vanishes on $\widetilde{A}$ and asymptotes to a vector $\zeta_A$ that defines an asymptotic symmetry for any asymptotically AdS geometry.

Now consider more generally asymptotically AdS spacetimes in the vicinity of pure AdS. For any such spacetime, we can choose a gauge for which the extremal surface $\widetilde{A}$ has the same coordinate location as in pure AdS and for which the Killing equation ${\cal L}_{\xi_A} g= 0$ for the original vector $\xi_A$ still holds at $\widetilde{A}$ \cite{Hollands:2012sf,Lashkari:2015hha}. Away from $\widetilde{A}$, ${\cal L}_{\xi_A} g$ will be non-zero, but we can think of it as defining a flow on the space of asymptotically AdS metrics.

Suppose for the moment that we restrict to metrics satisfying the field equations for some gravitational Lagrangian (in our case, this will be Einstein gravity minimally coupled to matter fields). Thinking of this gravitational theory as a system described by Hamiltonian mechanics, it turns out that the flow defined by the Lie derivative ${\cal L}_{\xi_A}$ can be understood as the flow on phase space associated with some corresponding Hamiltonian function $H_{\xi_A}$ \cite{Hollands:2012sf,Lashkari:2016idm}. Then for any other perturbation to the metric and matter fields, the change in $H_{\xi_A}$ is given by\footnote{ This follows from the standard relation $d H_X = i_X \Omega = \Omega (X, \cdot)$ between a Hamiltonian vector field $X$ and the associated Hamiltonian $H_X$. }
\be
\label{Ham}
{d H_{\xi_A} \over d \pert} = \Omega({d g \over d \pert}, {\cal L}_{\xi_A} g)
\ee
where $\Omega$ is the symplectic form on the phase space for the theory (and we are using $g$ to denote both metric and matter fields). We can think of the Hamiltonian $H_{\xi_A}$ as an energy associated with the timelike vector $\xi_A$. It can be expressed as a boundary term, the difference between $S^{grav}_A$, the area of the surface $\widetilde{A}$ for the case of Einstein gravity minimally coupled to matter, and a quantity $E^{grav}_A$ that we can think of as a charge associated with the asymptotic symmetry $\zeta_A$ \cite{Hollands:2012sf,Lashkari:2016idm}. The symplectic form $\Omega$ can be expressed as an integral over the region $\Sigma_A$ of a form $\bs\omega(\delta g_1, \delta g_2)$ that is a local expression in the two metric / matter field perturbations and their first derivatives. Thus, we have
\be
\label{Ham2}
{d \over d \pert} (E^{grav}_A - S^{grav}_A) = \int_{\Sigma_A} \bs\omega(g, {d g \over d \pert}, {\cal L}_{\xi_A} g) \; .
\ee

Now, returning to the full space of asymptotically AdS geometries with matter fields, we do not expect (\ref{Ham2}) to hold, but the difference between the two sides must be some quantity that vanishes when the gravitational equations are satisfied. It turns out that this difference is given by an integral over $\Sigma_A$ of a simple local expression ${\cal G}_{grav}$ built from the tensors appearing in the field equations for metric and matter fields. In this way, we obtain our key geometrical identity \cite{Hollands:2012sf}
\be
\label{Ham3}
{d \over d \pert} (E^{grav}_A - S^{grav}_A) = \int_{\Sigma_A} \bs\omega(g, {d g \over d \pert}, {\cal L}_{\xi_A} g) + \int_{\Sigma_A} {\cal G} \; .
\ee
Mathematically, this is simply a consequence of Stokes theorem.

Applying our quantum identity (\ref{RE}) to some family $|\Psi(\pert) \rangle$ of states and taking a $\pert$ derivative, we find a similar-looking identity that holds in the CFT,
\be
\label{RE2}
{d \over d \pert}(\langle H_A\rangle -  S_A) = {d \over d \pert} S(\rho_A || \rho^{(0)}_A) \; .
\ee
The two identities (\ref{Ham3}) and (\ref{RE2}) will form the basis of our proof.

\subsubsection*{Basic idea of the proof}

Suppose that a family of geometries $M(\pert)$ described by the metric
\be
\label{Mepsilon}
g(\pert) = g^{(0)} + \pert\, \delta g^{(1)} + \pert^2 \delta g^{(2)} + \dots
\ee
correctly calculates via HRRT the entanglement entropy for some family $|\Psi(\pert) \rangle$ of states up to some order in $\pert$. Then $S_A$ in (\ref{RE2}) must agree with $S^{grav}_A$ in (\ref{Ham3}).

We can also show that the remaining terms on the left sides of (\ref{Ham3}) and (\ref{RE2}) must agree with each other: Applying (\ref{RE2}) to an infinitesimal ball of radius $R$, we find that the terms on the right hand side are subleading in powers of the radius, so we have that
\be
{d \over d \pert} (S^{grav}_A) = {d \over d \pert} (S_A) = {d \over d \pert} (\langle H_A\rangle) \; .
\ee
As shown in \cite{Faulkner:2013ica}, this implies that in Fefferman-Graham gauge where $\Delta g_{zz} = \Delta g_{z \mu} = 0$, the leading order asymptotic metric must be related to the CFT stress tensor as
\be
\label{holoDict}
\Delta g_{\mu \nu} = {16 \pi \over d \ell^{d-3}} \langle T_{\mu \nu} \rangle z^{d-2} + \dots \; ,
\ee
where the dots indicate terms at lower orders in $z$. Using this relation and the explicit forms for $E^{grav}_A$ and $H_A$ it then follows that $E_A$ in (\ref{RE2}) must agree with $\langle H_A \rangle$ in (\ref{Ham3}) \cite{Faulkner:2013ica}. Since the left sides of these equation have now been shown to agree, we have that for the geometries $M(\pert)$ and the states $\Psi(\pert)$,
\be
\label{Key}
{d \over d \pert} S(\rho_A || \rho^{(0)}_A) = \int_{\Sigma_A} \bs\omega (g, {d g \over d \pert}, {\cal L}_{\xi_A} g) + \int_{\Sigma_A} {\cal G} \; .
\ee
This central result is valid for any $\pert$; in order to proceed, we now consider terms at specific orders in $\pert$ expanded about $\pert = 0$.

\subsubsection*{First order}

Setting $\pert = 0$ in the expression (\ref{Key}), the left side vanishes since the relative entropy between a reference state and a perturbation to that state vanishes at first order in the perturbation. The first term on the right side also vanishes, since the ${\cal L}_{\xi_A} g = 0$ for the unperturbed AdS metric. Further, the matter fields do not enter the expressions at this order. Thus, only the last term ${\cal G}$ is nonzero, and the the explicit form of the equation becomes
\be
\label{intEE1}
\int_{\Sigma_A} \xi_A^a E_{ab}^{(1)} \bs\epsilon^b = 0 \,,
\ee
where $E_{ab}^{(1)}$ is the Einstein tensor linearized about AdS and $\bs\epsilon^a$ is a volume form. A simple argument \cite{Faulkner:2013ica} shows that this is possible for all balls $A$ if and only if $E_{ab}^{(1)}$ = 0, so we see that the first order Einstein equations must be satisfied for any geometry that reproduces the entanglement entropy of a perturbed CFT state.\footnote{ In practice, one first shows that the equations of motion in the boundary directions are satisfied, $E^{(1)}_{\mu\nu} = 0$. The vanishing of the remaining components $E^{(1)}_{z\mu}$ and $E^{(1)}_{zz}$ can be thought of as constraint equations and follow from standard general relativity arguments. We refer to \cite{Faulkner:2013ica} and references therein for details.} We can further argue that a unique such geometry exists for every state of every CFT: given the CFT stress-energy tensor expectation value, we can associate a unique metric by solving the linearized Einstein equations with boundary conditions determined by (\ref{holoDict}). Reversing the steps that lead to (\ref{intEE1}) then implies that the HRRT formula for this geometry correctly calculates the entanglement entropies.

\subsubsection*{Second order}

At second order, our master equation (\ref{Key}) becomes explicitly
\be
\label{Key2}
\delta^{(2)} S(\rho_A || \rho^{(0)}_A) \equiv \frac{d^2}{d\pert^2} \, S(\rho_A || \rho^{(0)}_A) \big{|}_{\pert=0}
 = \int_{\Sigma_A} \bs\omega(g^{(0)}, \delta g^{(1)}, {\cal L}_{\xi_A} \delta g^{(1)})  - \int_{\Sigma_A}2\, \xi^a_A\, E_{ab}^{(2)}\bs\epsilon^b \; .
\ee
On the left side, $\delta^{(2)}S(\rho_A|| \rho_A^{(0)})$ is quadratic in the first order perturbation to the state and defines the {\it quantum Fisher information metric} at the point $\rho_A^{(0)}$ in the space of density matrices. The first term on the right defines the {\it canonical energy}; in a gravitational theory, it is interpreted as the perturbative energy for graviton and matter fluctuations with energy defined with respect to the timelike Killing vector $\xi_A$.

The results of the previous section show that the metric perturbation $\delta g^{(1)}$ satisfies Einstein's equations and is therefore determined in terms of the CFT stress tensor via a boundary-to-bulk propagator. The tensor $E_{ab}^{(2)}$ is the second order Einstein tensor minus the matter stress-tensor sources, which includes terms quadratic in the first order metric and matter perturbations and terms linear in the second order metric perturbation.

In order to proceed, we now restrict to the class of states (\ref{PIstate0}) that we expect to be well-described by some classical spacetime in the case of a holographic theory. In this case, we can compute the left side of $(\ref{Key2})$ perturbatively in the sources $\lambda_\alpha$. At quadratic order in the source $\lambda_\alpha$ for some operator ${\cal O}_\alpha$, the result for $\delta^{(2)}S(\rho_A|| \rho_A^{(0)})$ in terms of $\lambda_\alpha$ depends only on the two-point function of ${\cal O}_\alpha$ in the CFT (and its analytic continuation). With some work, it is possible to show that the CFT result for the second order relative entropy can be reexpressed in gravitational language,\footnote{ This should not be too surprising: the two-point function of an operator ${\cal O}_\alpha$ in any CFT is the same as the two-point function of an operator of the same dimension in a holographic CFT. We know that the latter can be alternatively obtained via a gravity calculation. Thus, we can re-express the two-point function for any CFT in gravitational language.} making use of the same auxiliary anti-de Sitter space as computes the leading order entanglement entropies. We define a field $\phi_\alpha^{(1)}$ on $g^{(0)}$ for each sourced operator ${\cal O}_\alpha$ in (\ref{PIstate0}), such that $\phi_\alpha^{(1)}$ satisfies the linearized equations of motion on AdS appropriate to the dimension and spin of ${\cal O}_\alpha$, and has boundary behaviour determined by $\langle {\cal O}_\alpha \rangle$ as we would have in genuine examples of holography. Then we can show that
\be
\label{Key3}
\delta^{(2)}S(\rho_A|| \rho_A^{(0)}) = \frac{\widetilde{C}_T}{a^*} \int_{\Sigma_A} \bs\omega(g^{(0)}, \delta g^{(1)}, {\cal L}_{\xi_A} \delta g^{(1)}) \; .
\ee
Deriving this equation will be the main task of the paper and will be the content of \S\ref{sec3}.

Finally, assuming that $\widetilde{C}_T = a^*$, we can combine (\ref{Key3}) and (\ref{Key2}) to obtain
\be
\int_{\Sigma_A} \xi^a \, E_{ab}^{(2)} \, \bs\epsilon^b = 0 \,.
\ee
Since this must be true for the surface $\Sigma_A$ associated with an arbitrary ball in an arbitrary frame of reference, the same argument as in the first order case shows that
\be \label{eq:EFE}
E_{ab}^{(2)} \equiv (E_{ab}^{(2)})_{grav} -  \frac{1}{2}T_{ab}^{(2)} = 0 \; ,
\ee
where $(E_{ab}^{(2)})_{grav}$ is the second order Einstein tensor and $T_{ab}^{(2)}$ is the second order stress-energy tensor associated with the matter fields.
Thus, any geometry that correctly calculates via HRRT the entanglement entropy of a CFT state of the form \ref{PIstate0} up to second order must satisfy Einstein's equations up to second order with matter determined by the CFT one-point functions.

To show that such a geometry exists for any state of the form (\ref{PIstate0}) in any CFT, we simply take the solution of Einstein's equations with asymptotic fields determined by the CFT one-point functions. In this case, we can reverse our steps to show that $S_A = S^{grav}_A$ up to second order.

\subsubsection*{Outline}

In the remainder of the paper, we explain in more detail the geometrical identity (\ref{Ham3}) in section \S\ref{sec2} and derive the CFT result (\ref{Key3}) in \S\ref{sec3} to complete the proof. In \S\ref{sec:discussion}, we explain in more detail why our general result is necessary for the validity of the AdS/CFT correspondence, and suggest how the result may generalize to theories with $a^* \ne \widetilde{C}_T$ and to higher orders in the state perturbation.

\section{Geometrical identities} \label{sec2}

In this section, we describe the geometrical aspects of our calculation, explaining in more detail the key identity (\ref{Ham3}).

\subsection{The Hollands-Wald gauge}
\label{sec:HW}

First, consider pure Poincar\'e-AdS with some curvature length scale $\ell$; the metric is given explicitly in (\ref{AdS}). For a boundary ball $A$ of radius $R$ centered at $x=x_0$ on a time slice $t=t_0$, the corresponding extremal HRRT surface $\widetilde{A}$ in pure AdS is given by $\{t= t_0, z^2 + (x - x_0)^2 = R^2\}$. The domain of dependence (causal diamond) of the spatial region $\Sigma_A$ between $A$ and $\widetilde{A}$ (see figure \ref{fig:hyperbolic2.pdf}) represents a Rindler wedge of AdS; it admits a timelike Killing vector $\xi_A$ vanishing at $\widetilde{A}$ and given explicitly by
\be\label{defxi}
\xi_A  = - \frac{2\pi}{ R}  (t-t_0) [z \partial_z + (x^i-x^i_0) \partial_i ] + \frac{\pi}{R} [R^2 - z^2 - (t-t_0)^2 - (\vec{x}-\vec{x_0})^2] \, \partial_t \; .
\ee
At the AdS boundary, this approaches the vector field
\be\label{defzeta}
\zeta_A  = - \frac{2\pi}{ R}  (t-t_0) [(x^i-x^i_0) \partial_i ] + \frac{\pi}{R} [R^2 - (t-t_0)^2 - (\vec{x}-\vec{x_0})^2] \, \partial_t
\ee
which defines an asymptotic symmetry for asymptotically AdS spacetimes (see also figure \ref{fig:flow2.pdf}).

Next, consider a one parameter family of asymptotically AdS spacetimes $M(\pert)$ described by metric
\be
\label{gexp}
g(\pert) = g^{(0)} + \pert\, \delta g^{(1)} + \pert^2 \delta g^{(2)} + \dots
\ee
where $g^{(0)}$ is the unperturbed metric (\ref{AdS}). Let $\widetilde{A}(\pert)$ be the extremal surface associated with the same boundary ball $A$ but in the new spacetime. As shown in \cite{Hollands:2012sf, Lashkari:2015hha} it is possible to choose a gauge for $g(\pert)$ such that:
\begin{enumerate}
\item The coordinate location of $\widetilde{A}(\pert)$ is fixed.
\item The vector $\xi_A$ with the same coordinate description as in the original AdS continues to satisfy the Killing equation $\left(\cL_{\xi_A} g(\pert)\right)_{\widetilde{A}} =0$ at the extremal surface.
\end{enumerate}
Here, the first condition is equivalent to the vanishing of the trace of the extrinsic curvature on $\widetilde{A}$. We will refer to this choice as ``Hollands-Wald'' gauge. This gauge is closely related to Gaussian normal coordinates, though it only constrains the metric near $\widetilde{A}$.

We will mostly use these gauge conditions applied to the first-order metric perturbation $\delta g^{(1)}$. In this case we can write the two conditions more explicitly as \cite{Lashkari:2015hha}:
\begin{enumerate}
\item $ \left( \nabla^{(0)}_\alpha \delta g^\alpha_{\;\bar{\a}} - \frac12 \nabla^{(0)}_{\bar{\a}} \delta g^\alpha_{\;\alpha} \right)_{\widetilde{A}} =0$, where $\alpha$ and $\bar{\a}$ are indices of coordinates respectively tangent and normal to $\widetilde{A}$, and
\item $\left(\cL_{\xi_A} \delta g \right)_{\widetilde{A}} =0$ or more explicitly:  $\Big( \delta g^\alpha {}_{\bar{\alpha}}  = 0 \Big)_{\widetilde{A}}$ and  $ \Big(\delta g^{\bar{\alpha}} {}_{\bar{\beta}} - {1 \over 2} \delta^{\bar{\alpha}} {}_{\bar{\beta}} \delta g^{\bar{\gamma}} {}_{\bar{\gamma}}\Big)_{\widetilde{A}} =0$.
\end{enumerate}

\subsection{The gravitational phase space}

We now describe more explicitly the gravitational identity (\ref{Ham3}) used in our derivation. The identity is entirely off-shell (i.e. does not assume any gravitational constraints on the metric or other fields), but is motivated most easily via a discussion of the covariant phase space formulation of gravity. For a more in depth discussion, see \cite{Hollands:2012sf}.\footnote{ Some early references on the covariant phase space approach in quantum field theory include section 5 of \cite{Witten:1986qs} and \cite{Crnkovic:1987tz}.}

First, we define natural volume forms
\be
\bs\epsilon_{a_1 \cdots a_k}  =  \frac{1}{(d+1-k)!}\,\sqrt{-g}\,\varepsilon_{a_1 \cdots a_k b_{k+1}\cdots b_{d+1}}dx^{b_{k+1}}\wedge\cdots\wedge dx^{b_{d+1}}  \; .
\label{defeps}
 \ee
These are defined so that contracting with a set of $k$ orthogonal unit vectors gives the volume form in the subspace perpendicular to these vectors.

Now, consider any covariant gravitational Lagrangian density ${\cal L}\, \bs\epsilon$ (considered as a $(d+1)$-form). Under a variation of the metric and matter fields $\phi_\alpha$ we have
\be
\delta ({\cal L} \,\bs\epsilon) = - E^{ab} \delta g_{ab} \,\bs\epsilon - E^{\alpha}_\phi \delta \phi_{\alpha} \bs\epsilon + d \bs\theta(g, \delta g, \phi_\alpha, \delta \phi_\alpha)
\ee
where $E^{ab}=0$ give the gravitational equations of motion, $E^\alpha = 0$ are equations of motion associated with matter fields $\phi_\alpha$, and $d \bs\theta$ is a total derivative term that we would normally encounter when integrating by parts to derive the Euler-Lagrange equations for this system. We assume for now that matter fields do not couple to curvatures. However, they do couple to the metric in the usual way, so the matter fields appear in $E^{ab}$  via the stress-energy tensor
\beq
E^{ab} \equiv E^{ab}_{grav}(g) - \frac{1}{2} \, T^{ab}(g,\phi_\alpha)\,.
\eeq
For example, in Einstein gravity coupled to a scalar field $E^{ab}_{grav}$ would be the Einstein tensor, and $T^{ab}$ would be the scalar stress energy tensor. In the following paragraphs, we focus on the gravitational part of the Lagrangian for ease of notation; the matter contributions to the various expressions are derived analogously and we will provide explicit expressions later on.

In the covariant phase space formulation of gravity, the symplectic form used in relating Hamiltonian functions to phase space flows is defined directly in terms of the quantity $\bs\theta$. The symplectic form applied to any two perturbations $(\delta g_1,\delta \phi_1)$ and $(\delta g_2,\delta \phi_2)$ is the integral over a Cauchy surface of a $d$-form $\bs\omega$ defined by
\be
\bs\omega(g ; \delta g_1, \delta \phi_1,\delta g_2,\delta \phi_2) = \delta_1 \bs\theta(g,\phi, \delta g_2,\delta \phi_2) - \delta_2 \bs\theta(g,\phi, \delta g_1,\delta \phi_1) \; .
\ee

Our basic identity (\ref{Ham3}) follows from the fact that for any metric $g$, vector field $X$, and metric perturbation $\delta g$, $\bs\omega$ applied to the perturbation $(\delta g, \delta \phi)$ and the Lie derivatives
\be
({\cal L}_X g)_{ab} = \nabla_a X_b + \nabla_b X_a \qquad {\cal L}_X \phi =  X^a \nabla_a \phi \; .
\ee
is a total derivative up to terms $\mathcal{G}$ that vanish when the equations of motion associated with ${\cal L}$ are satisfied:\footnote{ For readers familiar with the Wald formalism, $\bs\chi = \delta {\bf Q}_X  - X \cdot \bs\theta (g, \delta g)$, where ${\bf Q}_X$ is a form whose derivative gives the Noether current associated with the diffeomorphism generated by $X$; see \cite{Wald:1993nt} for a more detailed discussion.}
\be
\label{HWlocal}
\bs\omega(g;\delta g,\delta \phi,  {\cal L}_X g,{\cal L}_X \phi) = d \bs\chi(g,\phi ; \delta g, \delta \phi,X)  - \mathcal{G}(g,\phi ;\delta g,\delta \phi ,X) \; ,
\ee
A general physically motivated derivation can be found in \cite{Hollands:2012sf}, but for now, we require only the identity relevant to Einstein gravity with cosmological constant minimally coupled to matter, and in this case we can write all quantities explicitly and verify the identity by direct calculation. We have:
\bea
\label{eqn:grav_symplectic_form}
\bs\omega &=& \bs\omega_{grav} + \bs\omega_\phi \cr
\bs\omega_\phi \big(\phi; \delta_1 \phi , \delta_2 \phi \big) &=&  \bs\epsilon_a  \left( \delta_1\phi \, \nabla^a \delta_2\phi - \delta_2\phi \, \nabla^a \delta_1 \phi \right) \,,\cr
\bs\omega_{grav} \big(g; \gamma^1 , \gamma^2 \big) &=& \frac{1}{16 \pi}
\bs\epsilon_a P^{abcdef} \Big( \gamma^2_{bc} \nabla_d \gamma^1_{ef} - \gamma^1_{bc} \nabla_d \gamma^2_{ef} \Big)\cr
P^{abcdef} &=& g^{ae}g^{bf}g^{cd} - \frac12 g^{ad}g^{be}g^{cf} -\frac12 g^{ab}g^{cd}g^{ef} -\frac12 g^{ae}g^{bc}g^{fd} + \frac12 g^{ad}g^{bc}g^{ef} \; \cr
\bs\chi(g; \delta g, X) &=& \frac{1}{16 \pi} \bs\epsilon_{ab} \Big( \delta g^{ac} \nabla_c X^b
-\frac12 \delta g_c {}^c \nabla^a X^b +X^c \nabla^b \delta g^a {}_c - X^b \nabla_c \delta g^{ac} + X^b \nabla^a \delta g_c {}^c \Big) \cr
&& \qquad + \bs\epsilon_{ab} X^b \delta \phi \nabla^a \phi \label{defchi}\cr
\mathcal{G}(g; {dg \over d \pert},X) &=&  X^c \bs\epsilon_c E^{ab} \partial_\pert g_{ab} +  X^c \bs\epsilon_c E^{\phi} \partial_\pert \phi   -  2 X^a \partial_\pert \Big( E_{ab} \,\bs\epsilon^b  \Big) \cr
E_\phi &=&   \left( \nabla_a \nabla^a  - m^2 \right) \phi \,. \cr
E_{ab} &=&  \frac1{16 \pi} \Big( R_{ab} - \frac12 g_{ab} R  +  g_{ab} \Lambda \Big) -\frac{1}{2} \,T_{ab}(\phi) \; . \cr
T_{ab} &=& \left[  \na_a \phi \na_b \phi - \frac12 g_{ab} \left( \na_c \phi \na^c \phi + m^2 \phi^2 \right) \right]\,,
\eea
Here, $\nabla$ is the covariant derivative with respect to the background metric $g(\pert=0)$. We note that the terms involving $\Lambda$ cancel out between the two terms in $\mathcal{G}_{grav}$ so $\Lambda$ does not appear anywhere in the identity (\ref{HWlocal}).

The geometrical identity (\ref{Ham3}) follows by applying (\ref{HWlocal}) to the case where $g$ is the metric of some spacetime $M(\pert)$ in our one-parameter family, $\delta g = \partial_\pert g$, and $X = \xi_A$ and integrating over a spatial surface $\Sigma_A$ bounded by $A$ and $\widetilde{A}$. Using Stokes' theorem, (\ref{HWlocal}) then gives
\be
\label{HWlocal1}
\int_A \bs\chi - \int_{\widetilde{A}} \bs\chi = \int_{\Sigma_A} \bs\omega + \int_{\Sigma_A} \mathcal{G}  \; .
\ee
At the surface $\widetilde{A}$, $\xi_A$ vanishes and we can show that \cite{Iyer:1994ys,Faulkner:2013ica}
\be
\int_{\widetilde{A}} \bs\chi = {1 \over 4} {d \over d \pert} {\rm Area}(\widetilde{A}) \equiv  {d \over d \pert} S^{grav}_A
\ee
while at the AdS boundary, we have \cite{Faulkner:2013ica}
\be
\int_A \bs\chi =  {d \ell^{d-3} \over 16 \pi} \, {d \over d \pert}\int_A \zeta_A^\mu\, g_{\mu \nu}^{(d-2)} \bs\epsilon^\nu \equiv {d \over d \pert} E^{grav}_A
\ee
where $g^{(d-2)}_{ab}$ are the terms in the asymptotic metric at order $z^{d-2}$ in the Fefferman-Graham expansion. We note that both boundary terms are purely geometrical, depending only on the metric

Combining these results, we can rewrite \eqref{HWlocal1} to obtain our key geometrical identity
\be
\label{Ham3b}
{d \over d \pert} (E^{grav}_A - S^{grav}_A) = \int_{\Sigma_A} \left\{ \bs\omega_{grav}(g ; {d g \over d \pert}, {\cal L}_{\xi_A} g)  + \bs\omega_{\phi}(g ; {d \phi \over d \pert}, {\cal L}_{\xi_A} \phi) \right\}+ \int_{\Sigma_A} \mathcal{G}(g,\phi ; {d g \over d \pert}, {d \phi \over d \pert}, \xi_A ) \; ,
\ee
This is the most general identity that we will consider in this work, but further generalizations (for instance to more, or non-minimally coupled matter fields) are straightforward.

As we discussed in the introduction, the identity (\ref{Ham3b}) may be interpreted as the statement that the vector field $\xi_A$ defines a Hamiltonian flow on the space of metrics on $\Sigma_A$ and that the variation of the corresponding Hamiltonian can be written as the boundary term on the left hand side. For a more detailed discussion of this Hamiltonian and its relation to CFT relative entropy in the context of AdS/CFT, see \cite{Lashkari:2016idm}.

For our derivation, we require a perturbative version of the identity (\ref{Ham3b}), obtained by taking a derivative and evaluating the resulting equation at $\pert=0$. When the first order perturbations satisfy the linearized equations of motion, this gives
\begin{align}
\label{eqn:grav_result}
&(E^{grav}_A - S^{grav}_A)^{(2)} =
\int_{\Sigma_A} \Bigg\{ \bs\omega_{grav} \Big( \delta g^{(1)} ,\cL_{\xi_A} \delta g^{(1)}  \Big) + \bs\omega_{\phi} \Big( \delta\phi^{(1)}  ,\cL_{\xi_A} \delta\phi^{(1)}   \Big)
- 2 \xi_A^a \left( E_{ab}^{(2)}\right) \bs\epsilon^b
 \Bigg\},
\end{align}
where $E_{ab}^{(2)}$ represents Einstein's equations at second order in $\pert$:
\be
E_{ab}^{(2)} = \Big( \frac{\delta^2 (E_{ab})_{grav}}{\delta g_{cd} \delta g_{ef}} \delta g^{(1)}_{cd} \delta g^{(1)}_{ef} + \frac{\delta (E_{ab})_{grav}}{\delta g_{cd} } \delta g^{(2)}_{cd}  \Big) + \frac{\delta^2 (E_{ab})_\phi}{\delta \phi^2} \delta \phi^{(1)} \, \delta \phi^{(1)}\,.
\ee
We emphasize that this result is valid assuming that the metric perturbations are taken to satisfy the Hollands-Wald gauge conditions described in \S\ref{sec:HW}. In the following subsection, we further analyze the expression \eqref{eqn:grav_result} and give an analogous expression for generic metric perturbations that do not satisfy the Hollands-Wald gauge conditions.

\subsection{Gauge dependence of $\bs\omega$}
\label{sec:gauge_omega}

As we described in the introduction, if we assume that the spacetime $M(\pert)$ correctly calculates the ball entanglement entropies for some CFT state via the HRRT formula, then the left side of (\ref{eqn:grav_result}) must equal the CFT relative entropy at second order for this state relative to the vacuum state. The next step in our derivation is to compare the direct CFT calculation of relative entropy for states of the form (\ref{PIstate0}) to the right side of (\ref{eqn:grav_result}).

The result (\ref{eqn:grav_result}) is valid assuming that the metric perturbations satisfy the Hollands-Wald gauge; however, in comparing with the direct CFT calculation, it will be useful to express the results in a more general gauge, following \cite{Lashkari:2015hha}.

Consider then a general first-order metric perturbation $h$. By some infinitesimal coordinate transformation, $x^a \to x^a + V^a$ we can transform $h$ into a perturbation $\gamma$ satisfying the Hollands-Wald gauge condition. The two gauge-equivalent metric perturbations are related by
\be
\label{hTrans}
\gamma = h + \cL_{V} g
\ee
where $V$ is constrained by the requirement that $\gamma$ satisfy the Hollands-Wald conditions. The term involving $\bs\omega$ in the expression (\ref{eqn:grav_result}) can now be written for the metric perturbation $h$ in a general gauge as
\begin{align}
\bs\omega_{grav} \Big( \gamma ,\cL_{\xi_A} \gamma \Big) = \bs\omega_{grav} \Big( h ,\cL_{\xi_A} h \Big)
+ \bs\omega_{grav} \Big( h ,\cL_{\xi_A} \cL_{V} g  \Big)
+ \bs\omega_{grav} \Big( \cL_{V} g ,\cL_{\xi_A} \big( h + \cL_{V} g \big) \Big)
\label{omega_trans}
\end{align}
where we have used that $\omega$ is linear in each of the latter two arguments.

Starting from (\ref{HWlocal}), it follows that when one of the arguments of $\bs\omega_{grav}$ is pure gauge (i.e. equal to $\cL_X g$) and the other satisfies the linearized equations $E_{ab}^{(1)}(g,\delta g)=0$, the result is a total derivative:
\be
\bs\omega_{grav}(\delta g,\cL_X g) = d \bs\chi(\delta g,X)\\
\ee
with $\bs\chi$ defined in (\ref{defchi}). Thus, we can rewrite (\ref{omega_trans}) as
\begin{align}
\bs\omega_{grav} \Big( \gamma ,\cL_{\xi_A} \gamma \Big) = \bs\omega_{grav} \Big( h ,\cL_{\xi_A} h \Big)
+ d\bs\chi \Big( h ,[\xi_A,V]\Big)
- d\bs\chi \Big(\cL_{\xi_A} \big( h + \cL_{V} g \big) , V \Big) .
\end{align}
where the commutator is defined as
\be
[\xi_A,V]^a = \xi_A^b \partial_b V^a - V^b \partial_b \xi_A^a
\ee
and we have used that $\cL_{\xi_A} g = 0$ to write
\be
\cL_{\xi_A} \cL_V g = [\cL_{\xi_A} , \cL_V] g = \cL_{[\xi_A ,V]} g \; .
\ee
Making use of the fact that $h + \cL_{V} g$ is the metric perturbation in the Hollands-Wald gauge and therefore satisfies the Hollands-Wald gauge conditions, we can show (see Appendix \ref{sec:App3}) that $\bs\chi \big(\cL_{\xi_A} ( h + \cL_{V} g ) , V \big)$ vanishes when evaluated on $\widetilde{A}$. Choosing $V$ to vanish sufficiently quickly at the asymptotic boundary, this implies that
\begin{align}\label{eq:chivanish}
\int_{\Sigma_A} d\bs\chi \big(\cL_{\xi_A} ( h + \cL_{V} g ) , V \big) =0 \,.
\end{align}
Thus, we conclude finally that if $h$ is a general metric perturbation which can be transformed to the perturbation $\gamma$ in the Hollands-Wald gauge via a diffeomorphism associated with vector $V$, then
\begin{align}
\int_{\Sigma_A}  \bs\omega_{grav} \Big( \gamma ,\cL_{\xi_A}  \gamma \Big) &= \int_{\Sigma_A}  \bs\omega_{grav} \Big(h ,\cL_{\xi_A} h \Big)  + \int_{\Sigma_A} \bs\omega_{grav} \Big(h, \cL_{[\xi_A,V]} g \Big) \cr
&= \int_{\Sigma_A} \bs\omega_{grav} \Big(h ,\cL_{\xi_A} h \Big) + \int_{\widetilde{A}} \bs\chi \Big( h ,[\xi_A,V]\Big) \; ,
\label{omega_trans2}
\end{align}
where again we assume that $V$ vanishes sufficiently rapidly at the AdS boundary. The second term in this equation can be interpreted as effectively applying the change of coordinates away from Hollands-Wald gauge to $\xi_A$.

From the Hollands-Wald conditions for $\gamma$ listed at the end of \S\ref{sec:HW}, we can write explicit conditions that allow us to determine an appropriate $V$ corresponding to $h$. These are
\beqn \label{cons1}
\left(\nabla^{(0)}_{\a}\nabla^{(0),\a} V_{\bar{\a}} + \left[\nabla^{(0)}_{\a}, \nabla^{(0)}_{\bar{\a}}\right]V^{\a}\right)_{\widetilde{A}}&=& -\left(\nabla^{(0)}_\a{ h^\a}_{\bar{\a}}- \frac{1}{2} \nabla^{(0)}_{\bar{\a}}{ h^\a}_{\a}\right)_{\widetilde{A}}\nonumber\\
\cL_{V} g^{(0)}|_{\widetilde{A}} &=&- \cL_{\xi_A}h |_{\widetilde{A}} .
\eeqn
The first of these equations is essentially the inhomogenous Laplace equation on the RT surface with the right hand side acting as a source, and can be solved straightforwardly in terms of the appropriate Green's function. The second equation then determines the first derivative of $V$ away from the RT surface. We refer the reader to \cite{Lashkari:2015hha} for additional details.

We can now write a general expression for the second order gravitational result \eqref{eqn:grav_result} without assuming that the perturbation is in Hollands-Wald gauge. Instead, we require the knowledge of the vector $V$, which performs such a gauge transformation and write
\begin{align}
&(E^{grav}_A - S^{grav}_A)^{(2)} \label{eqn:grav_result2}\\
&\qquad =
\int_{\Sigma_A} \Bigg\{ \bs\omega_{grav} \Big( h ,\cL_{\xi_A}h  \Big) + \bs\omega_{\phi} \Big( \delta\phi ,\cL_{\xi_A} \delta  \phi\Big)
- 2 \xi_A^a \left( E_{ab}^{(2)}\right) \bs\epsilon^b
 \Bigg\} + \int_{\widetilde{A}} \bs\chi \Big( h , [\xi_A,V] \Big)
, \nonumber
\end{align}
where we also re-instated the scalar field dependence for completeness.
This is our final result for the main gravitational identity used in our derivation. The goal of the next section is to compute the second order relative entropy in a general CFT and write it in a form that very closely resembles \eqref{eqn:grav_result2}.

\section{Relative Entropy from the CFT} \label{sec3}

In this section, we will perform a direct CFT calculation of the relative entropy comparing the reduced density matrix for a ball-shaped region in excited states of the form (\ref{PIstate0}) with the reduced density matrix for the same region in the vacuum state. We will show that at second order in the state deformation, the result can be expressed in terms of an auxiliary asymptotically AdS spacetime $M(\pert)$ with matter fields $\phi_\alpha$ corresponding to each operator ${\cal O}_\alpha$ in the definition of the state. Explicitly, we will find
\be
\label{CFTresult1}
\delta^{(2)}S(\rho_A|| \rho_A^{(0)}) = \frac{\widetilde{C}_T}{a^*}\int_{\Sigma_A} \bs\omega_{grav}(g^{(0)}, \delta g^{(1)}, {\cal L}_{\xi_A} \delta g^{(1)}) + \int_{\Sigma_A} \bs\omega_{\phi_\alpha}( \delta \phi_\alpha^{(1)}, {\cal L}_{\xi_A} \delta \phi_\alpha^{(1)})
\ee
where $g^{(0)}$ represents the auxiliary AdS space (\ref{AdS}) that correctly calculates the vacuum entanglement entropies for ball-shaped regions via HRRT, $\delta g^{(1)}$ is a metric perturbation in Hollands-Wald gauge solving linearized Einstein equations on this spacetime with boundary behavior determined in terms of the CFT stress-energy tensor by (\ref{holoDict}), and $\phi_\alpha$ are fields on $g^{(0)}$ solving linearized equations for scalar fields with mass given in terms of the dimension of the corresponding operator by $m^2_\alpha \ell^2 = \Delta_\alpha (\Delta_\alpha - d)$ and with asymptotic behavior determined by
\be
\label{phi_asympt}
\phi_\alpha(z,x) =  \frac{\langle {\cal O}_\alpha(x) \rangle}{2\Delta-d} \, \left( z^{\Delta_\alpha} + \ldots \right) \; ,
\ee
where the dots indicate higher orders in $z$.

We begin in \S\ref{sec:EuclPI} by motivating our choice of states. In \S\ref{sec:2ptfcs}, we show the formula for relative entropy at second order in a source $\lambda_\alpha$ used to define the state depends only on the analytically continued CFT two-point function for the corresponding operator ${\cal O}_\alpha$. In \S\ref{sec3.2}, we derive the result (\ref{CFTresult1}) for the simpler case of scalar deformations in some detail. The important case of stress tensor deformations will be treated in \S\ref{sec3.4}, where much of the discussion from the scalar case will carry over directly.

\subsection{Classical bulk states from the Euclidean path integral with sources}
\label{sec:EuclPI}

In the context of holographic CFTs, we expect that only a subset of states are dual to spacetimes with a good classical description; more generally we could have states representing superpositions of classical spacetimes or something non-geometrical. Thus, to uncover Einstein's equations from CFT physics, we expect that is important to focus on a particular class of excited states.

The states we consider are conveniently described using the Euclidean path integral. We recall that the vacuum state $|0\rangle$ of a CFT can be constructed by performing the path-integral of the CFT on the lower half Euclidean space $x_0^E < 0$
\beq
\langle \varphi^{(0)}(\vx) | 0 \rangle = \int^{\varphi(x^E_0= 0, \vx) = \varphi^{(0)}(\vx)} [D\varphi]\, e^{-\int_{-\infty}^0dx_0^E\int d^{d-1}\vx\,\cL_{CFT}[\varphi]},
\eeq
where $x_0^E$ is Euclidean time, $\vx$ denotes spatial coordinates, and $\varphi$ collectively denotes all the fields in the CFT which are integrated over in the path integral. The general states that we consider are states $|\psi_{\lambda}(\pert)\rangle$ obtained similarly from a Euclidean path-integral, but with the action deformed by turning on Euclidean sources $\lambda_\alpha$ for CFT primary operators $\mathcal{O}_\alpha$
\beq \label{Estate}
\langle \varphi^{(0)}(\vx) | \psi_{\lambda}(\pert) \rangle = \int^{\varphi(x^E_0= 0, \vx) = \varphi^{(0)}(\vx)} [D\varphi] \,e^{-\int_{-\infty}^0dx_0^E\int d^{d-1}\vx\,\left(\cL_{CFT}[\varphi]+\lambda_\alpha(x;\pert) \mathcal{O}_\alpha(x)\right)}.
\eeq
where $\lambda_{\alpha}(x;\pert) = \pert \lambda_{\alpha}(x) + O(\pert^2)$. Here, we need to take the sources to vanish sufficiently rapidly as $\tau \to 0$ in order to give finite energy states of the original theory \cite{Speranza:2016jwt}; a similar construction with sources that do not vanish defines states of some perturbed CFT.

Our states are motivated by the fact that in the context of AdS/CFT, they are expected to define excited states that roughly speaking correspond to coherent states of bulk fields \cite{Botta-Cantcheff:2015sav, Christodoulou:2016nej, MPRV}. In this context, we can restrict to primary operators ${\cal O}_\alpha$ dual to light fields in the bulk. We can think of the sources $\lambda_\alpha$ as giving boundary conditions that determine a Euclidean asymptotically AdS spacetime. Via an analytic continuation procedure described in \cite{Botta-Cantcheff:2015sav, Christodoulou:2016nej,MPRV}, this can be related to a Lorentzian spacetime that we associate with the state (\ref{Estate}).\footnote{ Specifically, we solve the Euclidean gravitational equations with boundary conditions given by our sources for $\tau < 0$ and by the complex conjugated sources for $\tau > 0$. This can be sliced at $\tau=0$ to read off Euclidean ``initial data,'' which is then analytically continued $(\phi_\alpha, \partial_\tau \phi_\alpha) \to (\phi_\alpha, i \partial_t \phi_\alpha)$ to give initial data for the associated Lorentzian spacetime.} To define the most general real Lorentzian spacetime, the Euclidean sources must be taken to be complex. At the linearized level, it is possible to understand the map from Euclidean sources to Lorentzian initial data very explicitly \cite{MPRV}; the work \cite{MPRV} suggests that a generic perturbation to the metric and fields can be approximated arbitrarily well by states of the form (\ref{Estate}) for an appropriate choice of the sources.

\subsection{Second order relative entropy from two-point functions}
\label{sec:2ptfcs}

Starting with states (\ref{Estate}) we would now like to compute the relative entropy $S(\rho_A || \rho_A^{(0)})$ to second order in the sources. From the basic definition
\beq \label{re1}
S(\rho_A || \rho_A^{(0)})  = \mathrm{Tr}\left(\rho_A \ln\,\rho_A - \rho_A\ln \rho_A^{(0)}\right),
\eeq
it is straightforward to show that the relative entropy vanishes at first order in the sources; the leading contribution is a quadratic form in the first order perturbation $\delta \rho = \partial_\pert \rho (\pert= 0)$. As we review in Appendix \ref{app0}, we can write this as
\be
\delta^{(2)}S(\rho_A || \rho_A^{(0)})  \equiv \frac{d^2}{d\pert^2}S(\rho_A || \rho_A^{(0)})\Big|_{\pert = 0} = 2\,F(\delta \rho, \delta \rho)
\ee
where $F$ is the quantum Fisher information metric which can be written explicitly as
\bea
\label{Fish}
F(\delta \rho_1, \delta \rho_2)
&=& -\frac{1}{2} \int_{-\infty}^{\infty} {ds \over 4 \sinh^2\left( \frac{s \pm i \epsilon}{2}\right)} \tr\left((\rho_A^{(0)})^{-1} \delta \rho_1 (\rho_A^{(0)})^{\pm {i s \over 2 \pi}} \delta \rho_2\, (\rho_A^{(0)})^{\mp{i s \over 2 \pi}}\right) \; .
\eea
This expression is symmetric under $\rho_1 \leftrightarrow \rho_2$.

We next determine $\delta \rho$. Starting from (\ref{Estate}), the reduced density matrix $\rho_A= \tr_{\bar{A}}|\psi_{\lambda}\rangle \langle \psi_{\lambda}|$ for a ball-shaped region $A$ can be obtained by defining a path integral over the entire Euclidean space, with complex conjugated sources $\lambda_\alpha(x_E^0,x) \equiv \lambda^*_\alpha(-x_E^0,x)$ for $x_E^0 > 0$, and a cut along the region $A$:
\be
\langle \varphi_- | \rho_A | \varphi_+ \rangle = {1 \over N_\lambda} \int^{\varphi(A^+) = \varphi_+}_{\varphi(A^-) = \varphi_-} [D\varphi] \,e^{-\int_{-\infty}^0dx_0^E\int d^{d-1}\vx\,\left(\cL_{CFT}[\varphi]+\lambda_\alpha(x;\pert) \mathcal{O}_\alpha(x)\right)}
\ee
For small $\lambda_\alpha$, one can use this path-integral description together with the fact that vacuum modular Hamiltonian for ball-shaped regions in CFTs is local to write down the following perturbative expansion for the reduced density matrix $\rho_A$:\footnote{ Here, we assume that ${\cal O}_\alpha$ has no expectation value in the vacuum state; otherwise we should replace ${\cal O}_\alpha$ with ${\cal O}_\alpha - \langle {\cal O}_\alpha \rangle {\bf 1}$.}
\beq
\label{deltarho}
\rho_A = \rho_A^{(0)} + \pert \int d^dx\,\lambda_\alpha(x) \,\rho_A^{(0)}\,\mathcal{O}_\alpha(x)
+ O(\pert^2) \; .
\eeq
In this expression, we recall that $\rho_A^{(0)} = e^{-H_A}/Z$, where $H_A$ is the vacuum modular Hamiltonian for $A$ defined in (\ref{modHam}). Euclidean evolution with $H_A/(2 \pi)$ defines a flow shown in figure \ref{fig:flow2.pdf} along an angular coordinate $\tau$ defined by
\be \label{eq:tauflow}
\partial_\tau = {1 \over R} \, x^0_E\, x^i \partial_i + {1 \over 2 R} (R^2 + (x^0_E)^2 - x^2) \partial_{x^0_E} \; .
\ee
In the second term in (\ref{deltarho}), the operator $\mathcal{O}_\alpha(x)$ is a Heisenberg operator defined with respect to this modular flow 
\be
\mathcal{O}_\alpha(x) \equiv \mathcal{O}_\alpha(\tau,\tilde{x}) \equiv e^{{\tau \over 2 \pi} H_A} {\cal O}_\alpha(0,\tilde{x}) e^{-{\tau \over 2 \pi} H_A} \frac{\Omega^{\Delta}(\tau,\tilde{x})}{ \Omega^{\Delta}(0,\tilde{x})} \; ,
\ee
where we take $0 \le \tau < 2 \pi$,  ${\cal O}(0,\tilde{x})$ is the corresponding operator localized on $A$, and $\tilde{x}$ represent the remaining coordinates. Because the modular Hamiltonian
generates a conformal transformation we must include conformal factors, which are defined in Appendix \ref{appA}.
\myfigtab{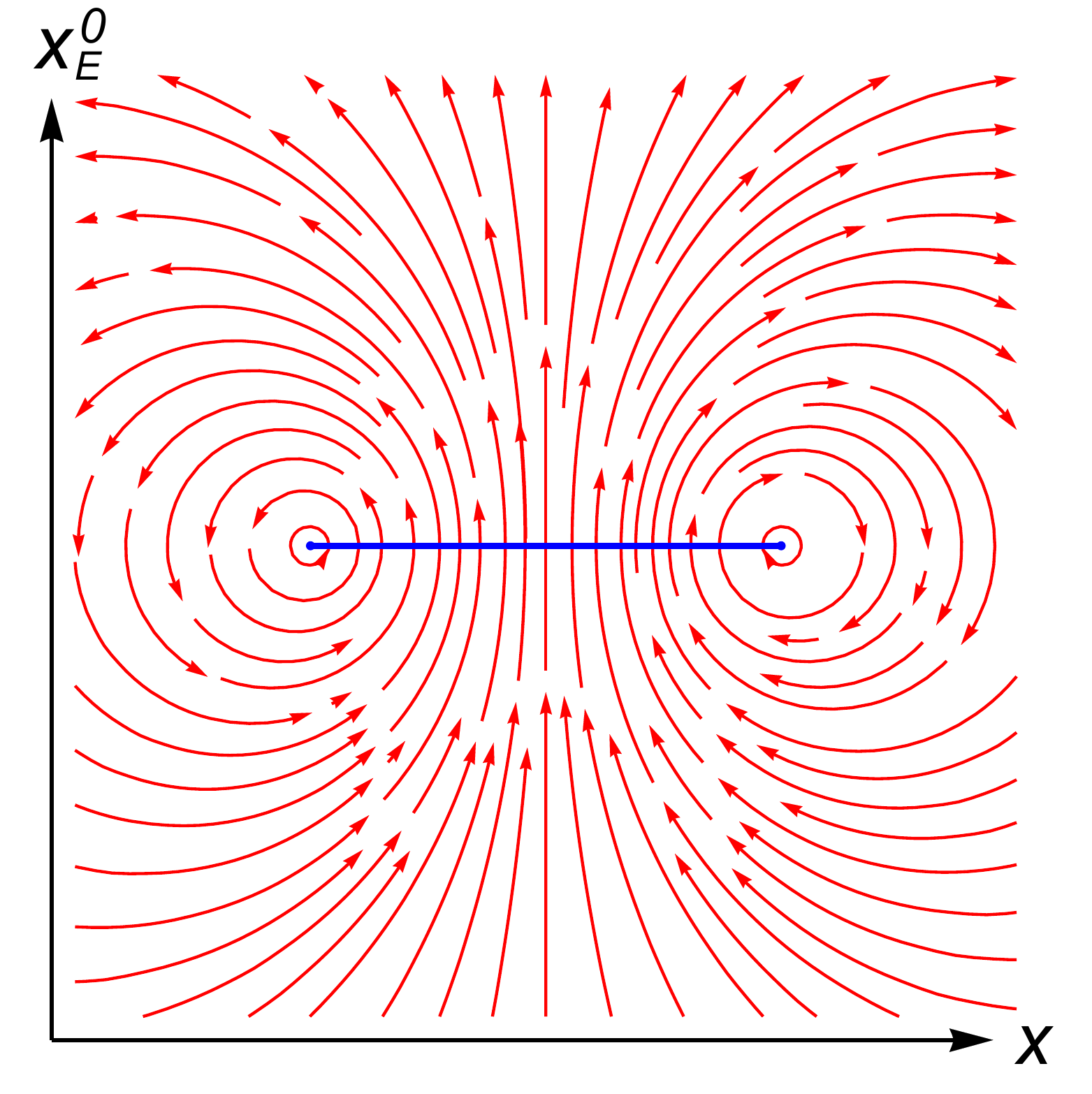}{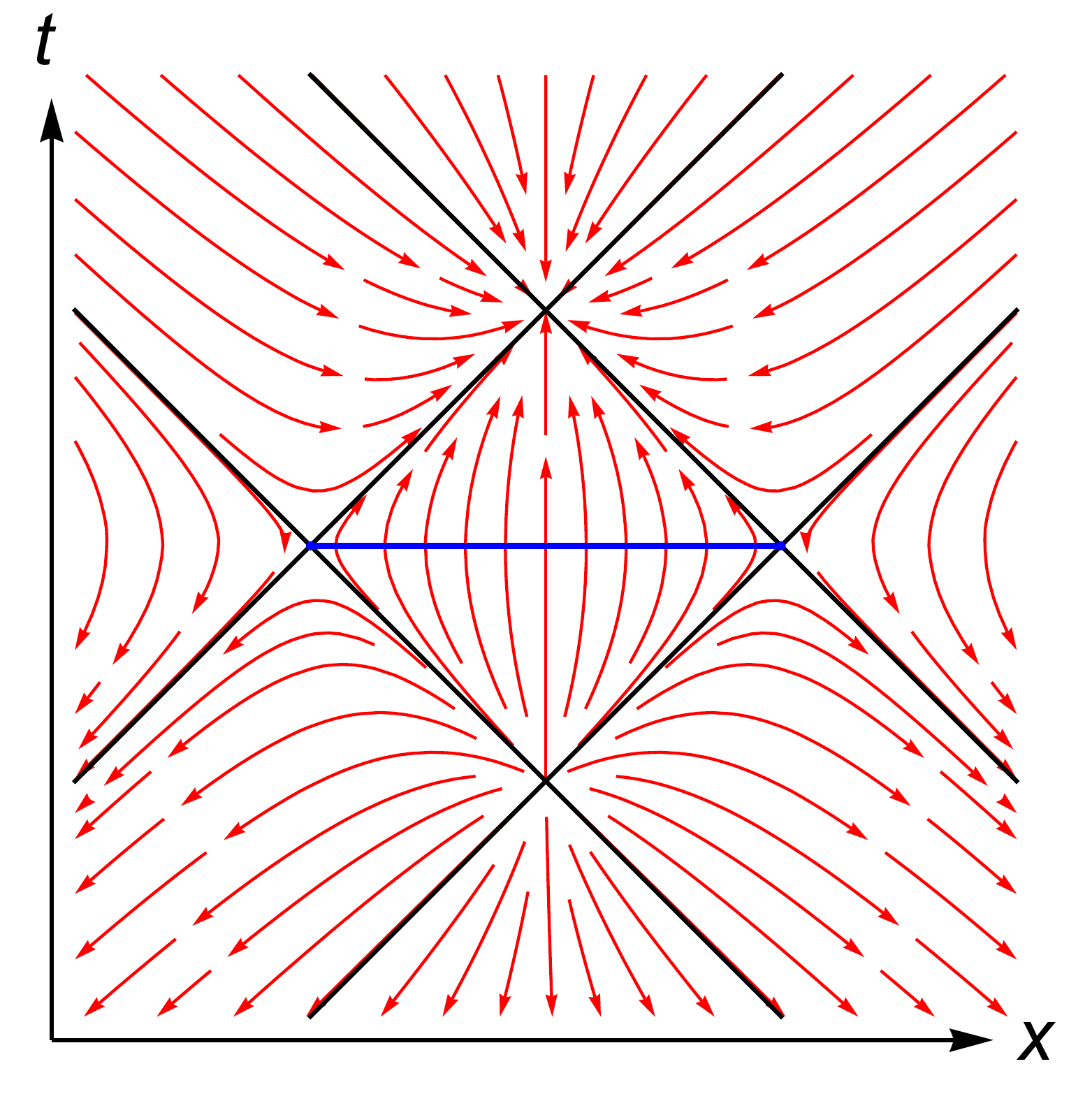}{6.5}{\small{\textsf{The flow generated by the modular Hamiltonian of the region $A$ (in blue). We start by working in Euclidean time $x^0_E$, depicting the vector field $\partial_\tau$ on the left. The calculation then introduces an imaginary shift in $\tau$, which results in a displacement of the operator insertion into real time $t$. The picture of the real time vector field $ \partial_s=\frac{1}{2\pi} \zeta_A$ is shown on the right with the causal structure indicated in gray. The two plotted planes can be thought of as orthogonal to each other and sewn together along the region $A$.}}}
We can now substitute the result (\ref{deltarho}) into (\ref{Fish}). The result is 
\beq \label{eq:fisherRes}
\begin{split}
& \delta^{(2)}S(\rho_A || \rho_A^{(0)}) \\
&\quad = 2\int d^d x_a \int d^d x_b \lambda_\alpha(x_a) \lambda_\beta(x_b) F(\rho_A^{(0)} {\cal O}_\alpha(x_a), \rho_A^{(0)} {\cal O}_\beta(x_b)) \\
&\quad = - \int d^d x_a d^d x_b \lambda_\alpha(x_a) \lambda_\beta(x_b)
\frac{\Omega^{\Delta}(\tau_b,\tilde{x}_b)}{ \Omega^{\Delta}(\tau_b+is,\tilde{x}_b)}
 \int_{-\infty}^{\infty} {ds \over 4 \sinh^2 \left({s + i \epsilon {\rm sgn}(\tau_a - \tau_b) \over 2 }\right)} \langle \mathcal{T} [{\cal O}_\alpha(\tau_a,\tilde{x}_a) {\cal O}_\beta(\tau_b + i s,\tilde{x}_b)] \rangle
\end{split}
\eeq
We refer the reader to Appendix \ref{app0} for a derivation of this identity. In the following we will consider a single operator for simplicity.

It is convenient at this stage to perform a conformal transformation from $\mathbb{R}^d$ to $S^1 \times \mathbb{H}^{d-1}$ (see \cite{Casini:2011kv} for further details). In these coordinates, $\tau$ becomes the coordinate on $S^1$, and the modular evolution by $H_A/(2 \pi)$ is simply translation around this circle. Then, equation \eqref{eq:fisherRes} becomes:
\beq\label{re2}
\delta^{(2)}S(\rho_A || \rho_A^{(0)}) =  - \int d\bs\mu \int_{-\infty}^{\infty} \frac{ds}{4\sinh^2\left(\frac{s+i\he\, \mathrm{sgn}(\tau_a- \tau_b)}{2}\right)} \left\langle \cO(\tau_a, Y_a) \cO(\tau_b+is, Y_b)\right\rangle ,
\eeq
where $Y$ are embedding space coordinates for $\mathbb{H}^{d-1}$ (see Appendix \ref{appA} for further details), the two point function appearing above is the analytically continued two-point function computed by the Euclidean path integral on $S^1 \times \mathbb{H}^{d-1}$ (which enforces $\tau$-ordering), and we have defined
\beq
\int d\bs\mu =\int_0^{2\pi} d\tau_a \int_{\mathbb{H}^{d-1}} dY_a\int_0^{2\pi} d\tau_b \int_{\mathbb{H}^{d-1}} dY_b \, \lambda(\tau_a, Y_a) \lambda(\tau_b, Y_b) \Omega^{\Delta -d }(\tau_a, Y_a) \Omega^{\Delta-d} (\tau_b, Y_b).
\eeq
Here, $\Omega(\tau, Y)$ is the conformal factor defined in Appendix \ref{appA}.

Our goal now is to express the result (\ref{re2}) in gravitational language, as in equation (\ref{Key3}). This result is expected for holographic theories \cite{Lashkari:2015hha}, where the right side is interpreted as the {\it canonical energy}, defined as the quadratic-order perturbative energy associated with the isometry $\xi_A$ in the Rindler wedge associated with $A$ (see figure \ref{fig:hyperbolic2.pdf}). But since (\ref{re2}) depends only on the CFT two-point functions, which have a universal structure, we shoud be able to express the result in the same way for general CFTs. In fact, for the case of scalar deformations which we will consider first, this was already shown in \cite{Faulkner:2014jva}.  In the next section we give a second, somewhat simpler argument, which we will then be able to generalize to the case of stress tensor deformations in the \S\ref{sec3.4}.

\subsection{Relative entropy for scalar deformations}\label{sec3.2}

In this section, we consider equation (\ref{re2}) in the case where the operators ${\cal O}_\alpha$ are scalar primary operators.

\subsection*{Asymptotic Symplectic Flux}

The main idea is that the CFT two-point function can be rewritten in terms of the symplectic flux $W_{\mathcal{D}(A)}$ at the asymptotic boundary of an \emph{auxiliary} AdS-Rindler wedge (namely the patch $\mathcal{D}(\Sigma_A)$ of an auxiliary AdS space, dual to $\mathcal{D}(A)$) as
\beq \label{sf1}
\begin{split}
&-\left\langle \cO(\tau, Y_a) \cO(is, Y_b)\right\rangle = W_{\mathcal{D}(A)}(K_E, K_R)\\
&\qquad\qquad \equiv  \lim_{r_0 \to \infty} \int_{r_B=r_0} ds_B\,dY_B\; \bs{\omega}_{\phi} \Big(K_E(r_B,i s_B,Y_B|\tau,Y_a), K_R(r_B,s_B,Y_B|s,Y_b)\Big)
\end{split}
\eeq
where $\bs{\omega}_{\phi}$ is the symplectic 2-form density corresponding to the bulk scalar field $\phi$ dual to the operator $\mathcal{O}$, and $K_E$ and $K_R$ are bulk-to-boundary propagators that we define below. The integral is over a constant radial slice at $r_B = r_0$ close to the asymptotic boundary of the Rindler wedge, or equivalently over $\mathcal{D}(A)$, namely the boundary domain of dependence of the region $A$ (see figure \ref{fig:fig31.pdf}). We use coordinates on the Rindler wedge for which the metric on this patch is
\beq
g^{(0)} = -(r_B^2-1)ds_B^2+\frac{dr_B^2}{(r_B^2-1)}+r_B^2 \, dY_B^2\, ;
\eeq
this takes the form of a special hyperbolic black hole in AdS$_{d+1}$,
where $Y_B$ parameterizes the horizon, i.e. the boundary of the Rindler wedge (see Appendix \ref{appA} for more details).
This hyperbolic black hole is familiar from the holographic context, but is used here purely to give a description of two-point functions as a symplectic flux in embedding space.
We emphasize that it is an auxiliary construct which is introduced without assuming anything about the CFT (in particular, the CFT does {\it not} have to be holographic).
\myfig{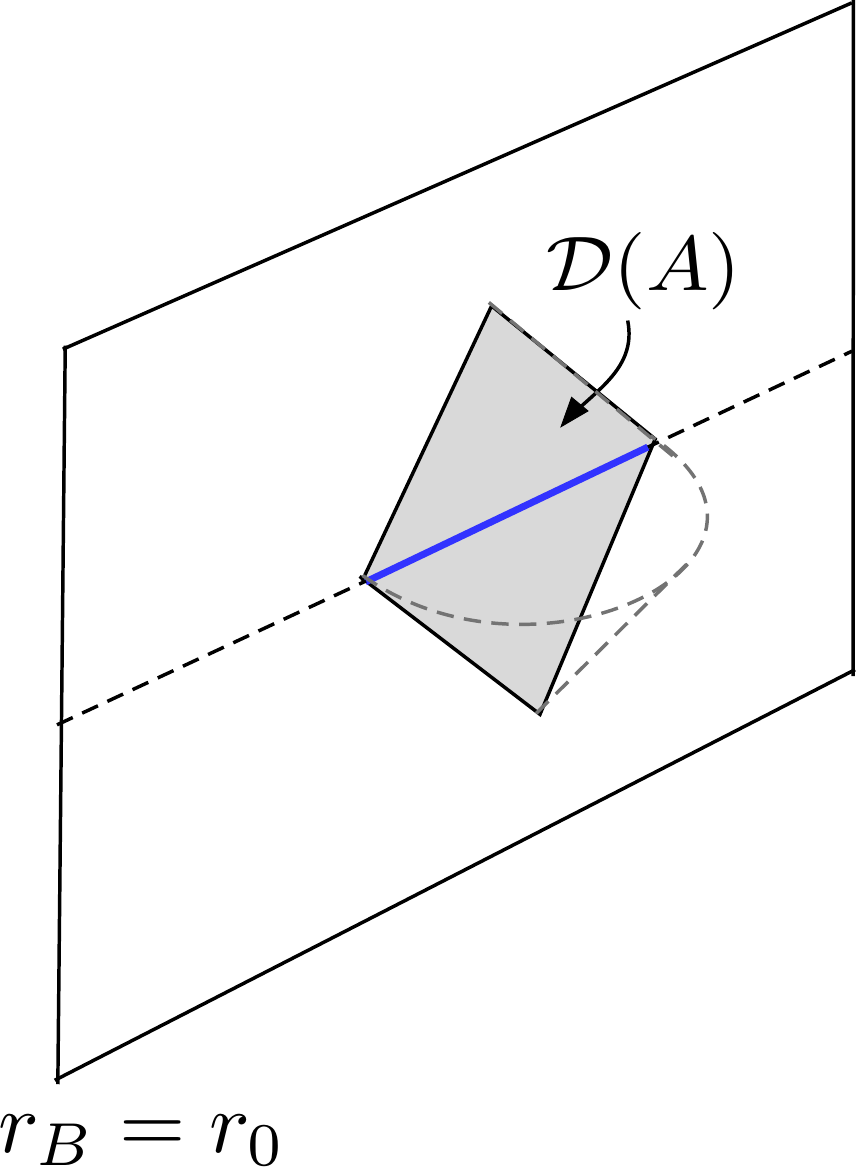}{5}{\small{\textsf{The two-point function can be written in terms of the symplectic flux integrated over the domain of dependence $\mathcal{D}(A)$ (the shaded region) of $A$ (blue line) on a radial slice of an auxiliary $AdS$ spacetime, close to the asymptotic boundary. }}}

In \eqref{sf1}, $K_E $ is the Euclidean bulk-to-boundary propagator sourced at Euclidean time $\tau$, but where the bulk point is analytically continued to real time
\beq
K_E (r_B,is_B,Y_B |  \tau, Y_a) = \frac{D_{\Delta}}{\left(-2 r_BY_B\cdot Y_a - 2\sqrt{r_B^2-1}\cos(\tau-is_B)\right)^{\Delta}},
\eeq
where $D_\Delta = \pi^{-d/2}\Gamma(\Delta) / \Gamma(\Delta-d/2)$.
Further, $K_R$ is the retarded bulk-to-boundary propagator
\beq \label{eq:Kret}
K_{R}(r_B,s_B,Y_B | s, Y_b) = i\,  \Theta(s_B-s_b) \Big(K_+(r_B,s_B, Y_B | s, Y_b) - K_{-}(r_B,s_B, Y_B | s, Y_b)\Big)
\eeq
where $K_{\pm}$ are the Wightman propagators defined as
\beqn \label{WtoE}
K_{\pm} (r_B,s_B,Y_B|s,Y_b) &=& \lim_{\varepsilon \to 0^+} K_{E} (r_B,is_B,Y_B|is\mp \varepsilon,Y_b)\nonumber\\
&=& \lim_{\varepsilon \to 0^+}  \frac{D_{\Delta}}{\left(-2 r_BY_B\cdot Y_b-2\sqrt{r_B^2-1}\cosh(s-s_B\pm i \varepsilon)\right)^{\Delta}}.
\eeqn

Let us now verify equation \eqref{sf1}. We note that the asymptotic flux for two solutions to the \emph{linearized} equations of motion $\delta_1\phi$ and $\delta_2\phi$ is given by
\beqn \label{eq:ScalarOmega}
W_{\mathcal{D}(A)}(\delta_1 \phi, \delta_2\phi) &=& \int_{r_B = r_0} \bs{\omega}_{\phi}(\delta_1 \phi, \delta_2\phi) \nonumber\\
&=&  \int_{r_B = r_0} \sqrt{-\gamma}\,n^M\left( \delta_1\phi \,\pa_M\delta_2\phi - \delta_2\phi\, \pa_M \delta_1\phi\right)\nonumber\\
&\simeq &  \int_{r_B =r_0} ds_B dY_B\,r_B^d\, \left( \delta_1\phi \,r_B\pa_{r_B}\delta_2\phi - \delta_2\phi\, r_B\pa_{r_B} \delta_1\phi\right)
\eeqn
where the limit $r_0\to \infty $ is implicitly assumed. Since $\delta_i\phi$ solve the linearized equations of motion, their asymptotic expansions take the following form:
\beq
\delta_{i}\phi(r_B,s_B,Y_B) \sim a_i(s_B,Y_B) r_B^{-d+\Delta}  + b_i(s_B, Y_B)r_B^{-\Delta},\qquad   (i=1,2)
\eeq
which then gives
\beq \label{sf2}
W_{\mathcal{D}(A)}(\delta_1 \phi, \delta_2\phi) = (2\Delta -d) \int ds_B dY_B \,\Big(a_1b_2 - a_2b_1\Big).
\eeq
Now from equation \eqref{sf1}, using the fact that $\delta_1\phi = K_E$ and $\delta_2\phi = K_R$, we have
\beq \label{ae1}
a_1 = 0, \;\;\;\; b_1 =  \frac{1}{(2\Delta-d)}\,G_E (is_B,Y_B|  \tau,Y_a )
\eeq
\beq\label{ae2}
a_2 = \delta^{d-1}(Y_B-Y_b) \delta(s_B-s),\;\;\; b_2 = \frac{1}{(2\Delta -d)}\,G_R(s_B, Y_B | s, Y_b)
\eeq
where $G_E$ is the Euclidean 2-point function on $S^1 \times \mathbb{H}^{d-1}$
\beq
\begin{split}
G_{E}(\tau_1,Y_1 | \tau_2, Y_2) \equiv \langle {\cal O}(\tau_1, Y_1){\cal O}(\tau_2, Y_2)\rangle  = \frac{2\Delta -d}{\pi^{d/2}} \frac{\Gamma(\Delta)}{\Gamma(\Delta-d/2)} \frac{1}{\left(-2Y_1\cdot Y_2 - 2\cos(\tau_1-\tau_2)\right)^{\Delta}}  \,
\end{split}
\eeq
analytically continued to real time in one of the arguments, while $G_R$ is the retarded two-point function on $\mathbb{R} \times \mathbb{H}^{d-1}$ (see Appendix \ref{app:propagators} for some more details). Further, $a_1= 0$ because the source point of the Euclidean bulk to boundary propagator $K_E$ is at the Euclidean time $\tau$, while the bulk point is at real time $s_B$. The coefficients required to match $b_1$ and $b_2$ to CFT Green's functions are borrowed from the standard AdS/CFT literature (see for instance \cite{Klebanov:1999tb}).

Combining equations \eqref{sf2}, \eqref{ae1} and \eqref{ae2}, we arrive at the desired result:
\beq
-G_{E}(\tau, Y_a | is, Y_b) =W_{\mathcal{D}(A)}(K_E, K_R) \,.
\eeq

\subsubsection*{Back to relative entropy}
Following the discussion above, we can now rewrite the relative entropy from equation \eqref{re2} as
\beq
\delta^{(2)} S(\rho_A||\rho_A^{(0)})_\phi  =  \int d\bs\mu \int_{-\infty}^{\infty} \frac{ds}{4\sinh^2\left(\frac{s+i\he\, \mathrm{sgn}(\tau_{a}-\tau_b)}{2}\right)} W_{\mathcal{D}(A)}(K_E, K_R)
\eeq
where the Euclidean bulk-to-boundary propagator $K_E$ is sourced at Euclidean time $\tau = (\tau_a - \tau_b)$, and the retarded propagator is sourced at real time $s$. The symplectic flux appearing in the above equation is evaluated at the asymptotic boundary $r_B \to \infty$.

The key point now is that the integrated symplectic 2-form density is conserved in the sense that it is invariant under deformations of the integration surface. This follows because the symplectic form is a closed differential form on spacetime \cite{Witten:1986qs, Crnkovic:1987tz}. In general, this property requires the fields to be on-shell; but the propagators appearing in our expression satisfy scalar field equations by definition. 

By deforming the integration surface, we can equally well evaluate the symplectic flux at the horizon of the hyperbolic black hole, namely at $r_B \to 1 $ (see figure \ref{fig:fig32.pdf}).
\myfig{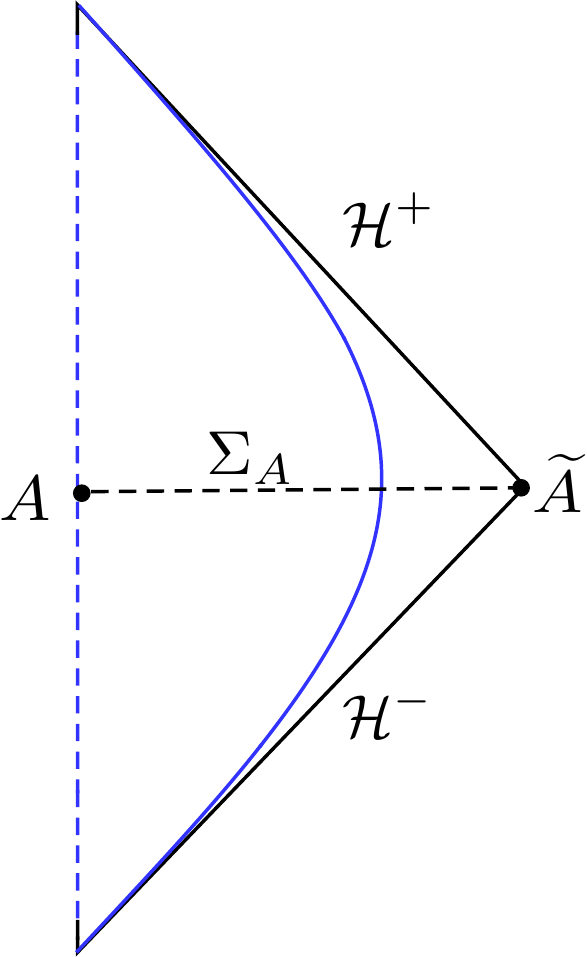}{4}{\small{\textsf{The asymptotic symplectic flux evaluated on $\mathcal{D}(A)$ at $r_B \to \infty$ (dashed blue line) is equal to the symplectic flux at the horizon $r_B \to 1$ (solid blue line).  }}}\\
We therefore write:
\beq
\delta^{(2)} S(\rho_A||\rho_A^{(0)})_\phi =  \lim_{r_0 \to 1} \int d\bs\mu \int_{-\infty}^{\infty} \frac{ds}{4\sinh^2\left(\frac{s+i\he\, \mathrm{sgn} (\tau_a - \tau_b)}{2}\right)} \int_{r_B=r_0}ds_B dY_B\,\bs{\omega}_{\phi}(K_E, K_R)
\eeq
Note that the retarded propagator $K_{R}(r_B,s_B,Y_B | s, Y_b)$ only has support over $s_B > s$, and additionally by causality is only non-zero when the bulk and the boundary points are time-like separated. So we obtain
\beq
\begin{split}
&\delta^{(2)} S(\rho_A||\rho_A^{(0)})_\phi  \\
 &\qquad = i\,\lim_{r_0 \to 1} \int d\bs\mu \int_{-\infty}^{\infty} \frac{ds}{4\sinh^2\left(\frac{s+i\he\, \mathrm{sgn} (\tau_a - \tau_b)}{2}\right)}\int_{ s_{B,*}^{\pm}}^{\infty} ds_B\int_{\mathbb{H}^{d-1}} dY_B\,\bs{\omega}_{\phi}(K_E,  K_+- K_-)_{r_B=r_0}
\end{split}
\eeq
where
$$s_{B,*}^{\pm} = s + \mathrm{ln}\left(\a\pm \sqrt{\a^2-1}\right),\;\;\; \a = - \frac{r_0 Y_B \cdot Y_b}{\sqrt{r_0^2-1}}$$
are the singularities associated with the two points inside the retarded bulk-to-boundary propagator becoming null (see figure \ref{fig:fig33.pdf}a).
\myfig{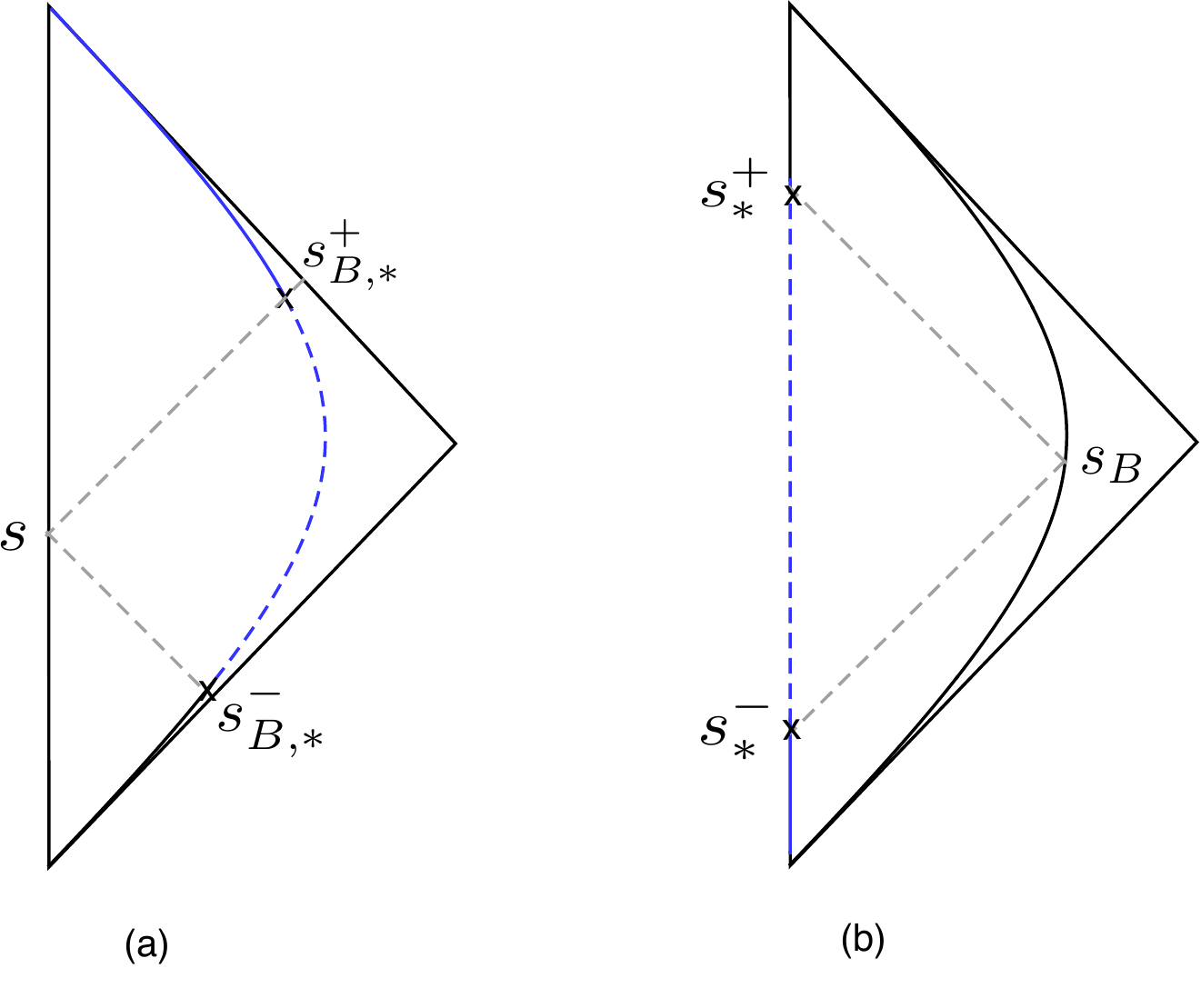}{8}{\small{\textsf{The null singularities of the Wightman propagators. (a) The $s_B$ integral is over the blue portion; the dashed blue line is the spacelike region which we can include for free. (b) The $s$ integral is over the blue portion, and once again we can include the dashed blue portion for free.}}}\\
The lower limit on the $s_B$ integral can be taken to be either $s_{B,*}^+$ or $s_{B,*}^-$, because as mentioned above the retarded propagator vanishes for spacelike separated points.
Exchanging the orders of the $s$ and $s_B$ integrals, we then obtain
\beq\label{re3}
\begin{split}
&\delta^{(2)} S(\rho_A||\rho_A^{(0)})_\phi   \\
&\qquad =i \int d\bs\mu  \int_{-\infty}^{\infty} ds_B\int_{\mathbb{H}^{d-1}} dY_B \int_{-\infty}^{s_*^{\pm}} \frac{ds}{4\sinh^2\left(\frac{s+i\he\,\mathrm{sgn}(\tau_a - \tau_b)}{2}\right)}\,\bs{\omega}_{\phi}(K_E,  K_+- K_-)_{r_B=r_0}
\end{split}
\eeq
where (see figure \ref{fig:fig33.pdf}b)
$$s_*^{\pm} = s_B - \mathrm{ln}\left(\a \mp \sqrt{\a^2-1}\right).$$
We now wish to study the second argument of $\bs \omega_\phi$ in \eqref{re3}, which means we need to perform the following $s$ integral in the limit where the bulk point approaches the horizon:
\beqn \label{I1}
I(r_B,s_B, Y_B|Y_b)  &=&i \int_{-\infty}^{s_*^{\pm}} \frac{ds}{4\sinh^2\left(\frac{s+i\he\, \mathrm{sgn} (\tau_a - \tau_b)}{2}\right)}\Big(K_+(r_B,s_B,Y_B|s,Y_b)- K_-(r_B,s_B,Y_B|s,Y_b)\Big)\nonumber\\
&=&i\,\lim_{\varepsilon \to 0^+} \int_{C_{\varepsilon} \cup C_{-\varepsilon}} \frac{ds}{4\sinh^2\left(\frac{s+i\he \, \mathrm{sgn}(\tau_a - \tau_b)}{2}\right)}K_E(r_B,is_B,Y_B|is ,Y_b)
\eeqn
where in the second line above, we have written the Wightman propagators $K_{\pm}$ in terms of Euclidean bulk-to-boundary propagators (following equation \eqref{WtoE}),\footnote{ We have also dropped an $O(\varepsilon)$ term. The order of limits is we send $\varepsilon \to 0^+$ at fixed $\he$, and then at the end of the calculation send $\he \to 0$.} and the contours above are defined as (see figure \ref{fig:fig34.pdf}a)
\beq
C_{\varepsilon} =   (-\infty+i\varepsilon , s_{*}^{-}+i\varepsilon),\;\;C_{-\varepsilon} =   ( s_{*}^{-}-i\varepsilon,-\infty-i\varepsilon).
\eeq
Note that by causality, we are allowed to simultaneously stretch these contours up to $s_*^+$ if we so desire. Next, it is convenient to use the KMS condition (i.e. the periodicity of $K_E$ in Euclidean time) to replace the contour $C_{-\varepsilon}$ with the equivalent contour (shown in figure \ref{fig:fig34.pdf}b)
\beq
C_{2\pi - \varepsilon} = (s_*^-+i(2\pi - \varepsilon), -\infty+i(2\pi - \varepsilon))
\eeq
This yields:
\beq\label{I2}
I(r_B,s_B, Y_B|Y_b) = i\int_{C_{\varepsilon} \cup C_{2\pi-\varepsilon}} \frac{ds}{4\sinh^2\left(\frac{s+i\he\, \mathrm{sgn} (\tau_a - \tau_b)}{2}\right)}K_E(r_B,is_B,Y_B|is ,Y_b)
\eeq
\myfigtab{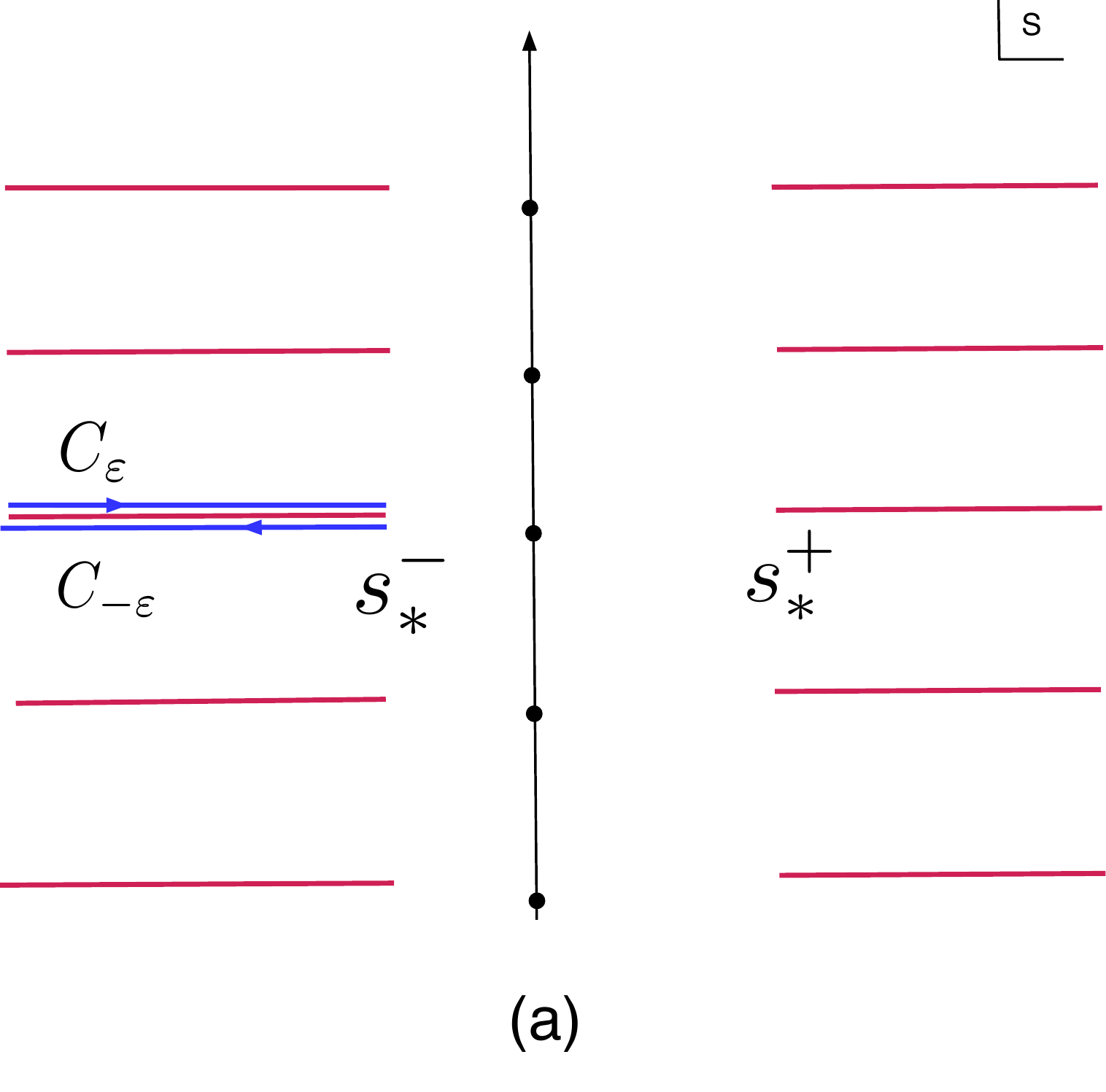}{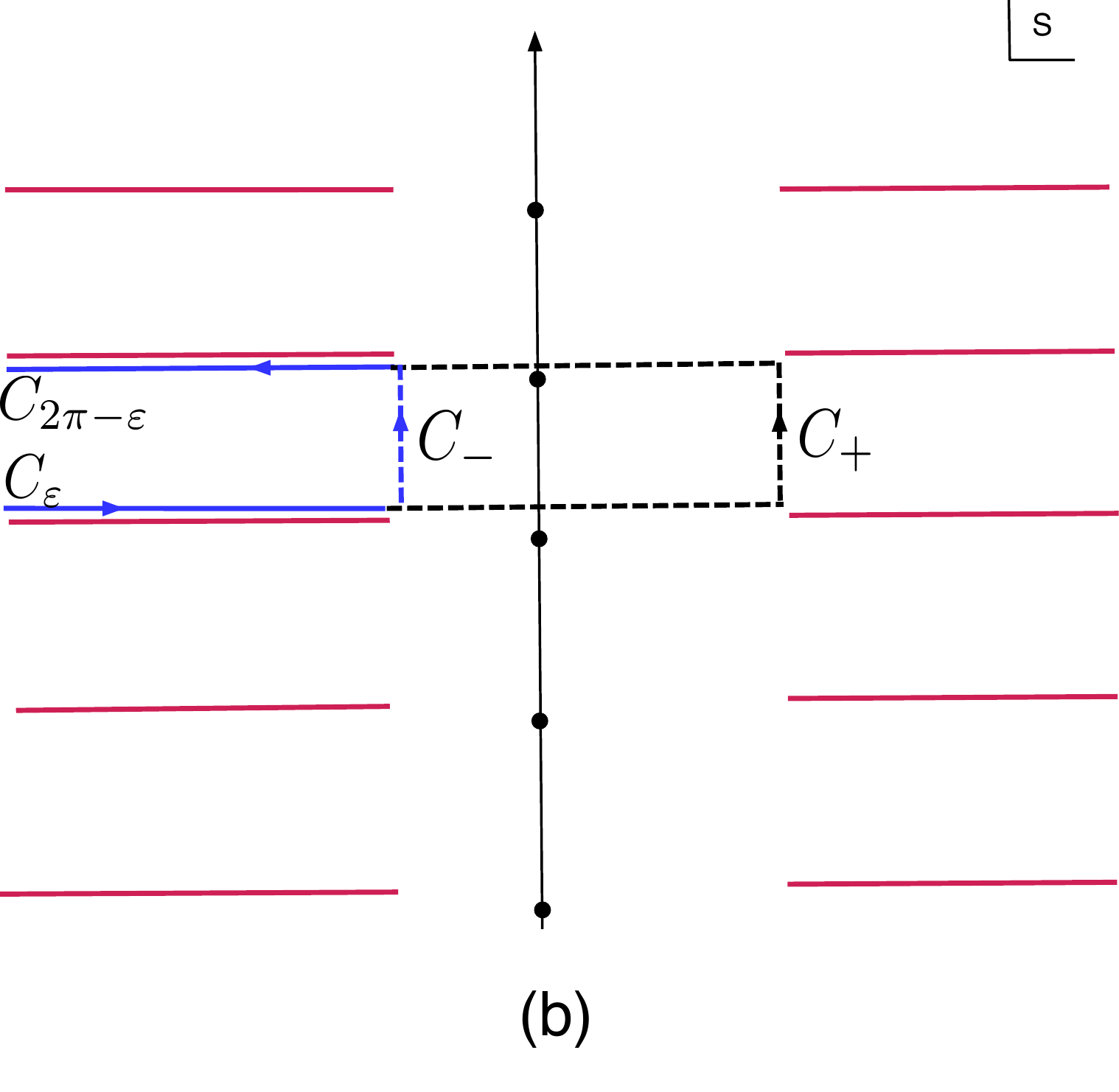}{6.2}{\small{\textsf{The analytic structure in the complex $s$-plane. The original integral is over the solid blue lines. On the future horizon, we complete the contour by including the dashed blue lines. On the past horizon, we complete it by including the dashed grey line at $s_*^-$.}}}
The analytic structure of the integrand in equation \eqref{I2} in the complex $s$ plane is displayed in figure \ref{fig:fig34.pdf}. The Euclidean bulk-to-boundary propagator has branch cuts  at $\mathrm{Re}(s) \leq s_*^-,\;\mathrm{Im}(s) = 2\pi \mathbb{Z}$ and $\mathrm{Re} (s) \geq s^+_*,\; \mathrm{Im}(s) = 2\pi \mathbb{Z}$ (denoted by red lines in figure \ref{fig:fig34.pdf}), while the $\sinh^2$ in the denominator has double poles at $s= -i\he\, \mathrm{sgn} (\tau_a- \tau_b)+2\pi i \mathbb{Z}$ (denoted by bold dots in figure \ref{fig:fig34.pdf}).

Now recall that we're taking the limit where the bulk point $(r_B, s_B,Y_B)$ is approaching the horizon of the hyperbolic black hole, $r_B \to 1$. Let us first consider the case where the bulk point approaches the past horizon, namely $r_B \to 1,s_B \to -\infty$ with $\sqrt{r_B^2-1}e^{-s_B} $ fixed, or in terms of light-cone coordinates
\beq
\ell_B^+ = \sqrt{r_B^2-1} \, e^{s_B},\qquad \ell_B^- = \sqrt{r_B^2-1} \, e^{-s_B}
\eeq
we have $\ell_B^+ \to 0, \;\ell_B^-$ fixed. In this case, we complete the contour in equation \eqref{I2} by including the vertical piece
\beq
C_- = (s_*^-+i\varepsilon, s^-_*+i(2\pi - \varepsilon)),
\eeq
denoted by the dashed blue line in figure \ref{fig:fig34.pdf}b. This is allowed because as we will show below, the integral along the contour $C_-$ vanishes as $O(\ell_B^+)$ in the limit that the bulk point approaches the past horizon.  But now using Cauchy's theorem, we see that the whole integral $I$ vanishes in this limit, because the integrand is analytic in the region enclosed by the full contour $C_{\varepsilon} \cup C_{-} \cup C_{2\pi - \varepsilon}$.\footnote{ The contribution from the contour at $s=-\infty$, namely $C_{-\infty}= (- \infty+i(2\pi -\varepsilon), -\infty + i\varepsilon)$ also vanishes because the integrand is exponentially suppressed as $s\to -\infty$.} So we conclude that
\beq
 \lim_{\ell_B^+\to 0}  I(\ell_B^+, \ell^-_B, Y_B|Y_b) = 0.
\eeq
This was more or less expected, because the retarded propagator vanishes on the past horizon.

On the other hand, as the bulk point approaches the future horizon, $\ell_B^- \to 0$ and $\ell_B^+$ fixed, we complete the contour of integration by stretching $C_{\varepsilon}$ and $ C_{2\pi-\varepsilon}$ up to $s_*^+$ (which is allowed by causality) and including the vertical piece
\beq
C_+ = (s_*^+ +i\varepsilon, s^+_*+i(2\pi - \varepsilon)),
\eeq
denoted by the dashed black line in figure \ref{fig:fig34.pdf}b. Once again, this is allowed because the integral along the contour $C_+$ vanishes as $O(\ell_B^-)$ as the bulk point approaches the future horizon, as we will show below.\footnote{ Note that while the contribution from $C_+$ vanishes on the future horizon, the contribution from $C_-$ does not vanish; this is why we have chosen to complete the contour by including $C_+$ instead of $C_-$.} But in contrast with the previous case, now the integral is non-zero because the completed contour includes  the double-pole at $s= i\he$ for $\tau_b>\tau_a$ or $s=i(2\pi - \he)$ for $\tau_a > \tau_b$. Therefore, we obtain using the residue theorem
\beq
\lim_{\ell_B^-\to 0}  I(\ell_B^+, \ell^-_B, Y_B|Y_b) = -2\pi  \pa_s K_E(\ell_B^+,0,Y_B | \,\mathrm{sgn}(\tau_a-\tau_b) \he, Y_b)
\eeq
where in the Euclidean propagator above it is understood that the bulk point has been analytically continued to the future horizon. From equation \eqref{re3}, we then have
\beq
\delta^{(2)}S(\rho_A||\rho_A^{(0)})_\phi = \int d\bs\mu \, W_{\mathcal{H}^+}\Big(K_E(\ell^+_B,0, Y_B| (\tau_{a}-\tau_b), Y_a), \mathcal{L}_{\xi_A}K_E(\ell^+_B, 0,Y_B|\, \mathrm{sgn}(\tau_a-\tau_b)\he, Y_b)\Big)
\eeq
where $W_{\mathcal{H}^+}$ is the symplectic flux evaluated on the future horizon, and additionally we have used translation invariance of $K_E$ along Rindler time to make the replacement
\beq
-2\pi\pa_s \to \mathcal{L}_{\xi_A}.
\eeq
Finally, we rescale the bulk coordinate $\ell^+_B \to \ell_B^+e^{-i\tau_b}$ and then rotate the new $\ell_B$ contour back to the positive real axis; it is easy to convince oneself that the $\he$ has the right sign so that this can be done without crossing any branch cuts (see figure \ref{fig:fig36.pdf}). This gives us the desired result
\beq \label{eq:scalarFinal}
\delta^{(2)}S(\rho_A||\rho_A^{(0)})_\phi = W_{\mathcal{H}^+}\left(\delta\phi,  \mathcal{L}_{\xi_A} \delta\phi\right)
 \equiv \int_{{\cal H}^+} \, \bs\omega_\phi(\delta \phi, \cL_{\xi_A} \delta \phi ) \; .
\eeq
where
\beq
 \delta\phi(\ell^+_B, Y_B) = \int d\tau dY\, \lambda(\tau, Y) \Omega^{d-\Delta}(\tau, Y) K_E( \ell^+_B, Y_B | \tau, Y) \,.
\eeq
The field $\delta \phi$ is the scalar field solving the source-free Lorentzian scalar field equations, whose analytic continuation has asymptotic behavior consistent with the sources $\lambda(\tau, Y)$, that is $\delta \phi_E \to_{r \to \infty} \lambda$. Alternatively \cite{MPRV}, it is exactly the on shell scalar field perturbation with asymptotic behavior determined by the CFT one point functions as (\ref{phi_asympt}).

The result (\ref{eq:scalarFinal}) is precisely the result derived in \cite{Faulkner:2014jva}. Since the symplectic 2-form density is conserved, one can freely move the symplectic flux from the future horizon to the region $\Sigma_A$ on a constant-time Cauchy slice enclosed between $A$ and the Ryu-Takayanagi surface $\widetilde{A}$. The deformed surface should also include a part at the AdS boundary that is the future domain of dependence of the region $A$, but this contribution vanishes due to the asymptotic fall-off of the fields at the AdS boundary (we need here that the Euclidean sources for $\delta \phi$ turn off at $\tau=0$). Note in particular that this derivation is much more efficient than the lengthy derivation of \cite{Faulkner:2014jva}, and relied very minimally on the exact form of the CFT correlators. Indeed this method can be used to derive similar results for QFT in curved space with an entangling cut at the bifurcation surface of a (conformal) Killing horizon where the appropriate correlator may not be known in closed form. We leave discussion of this to future work.

\myfig{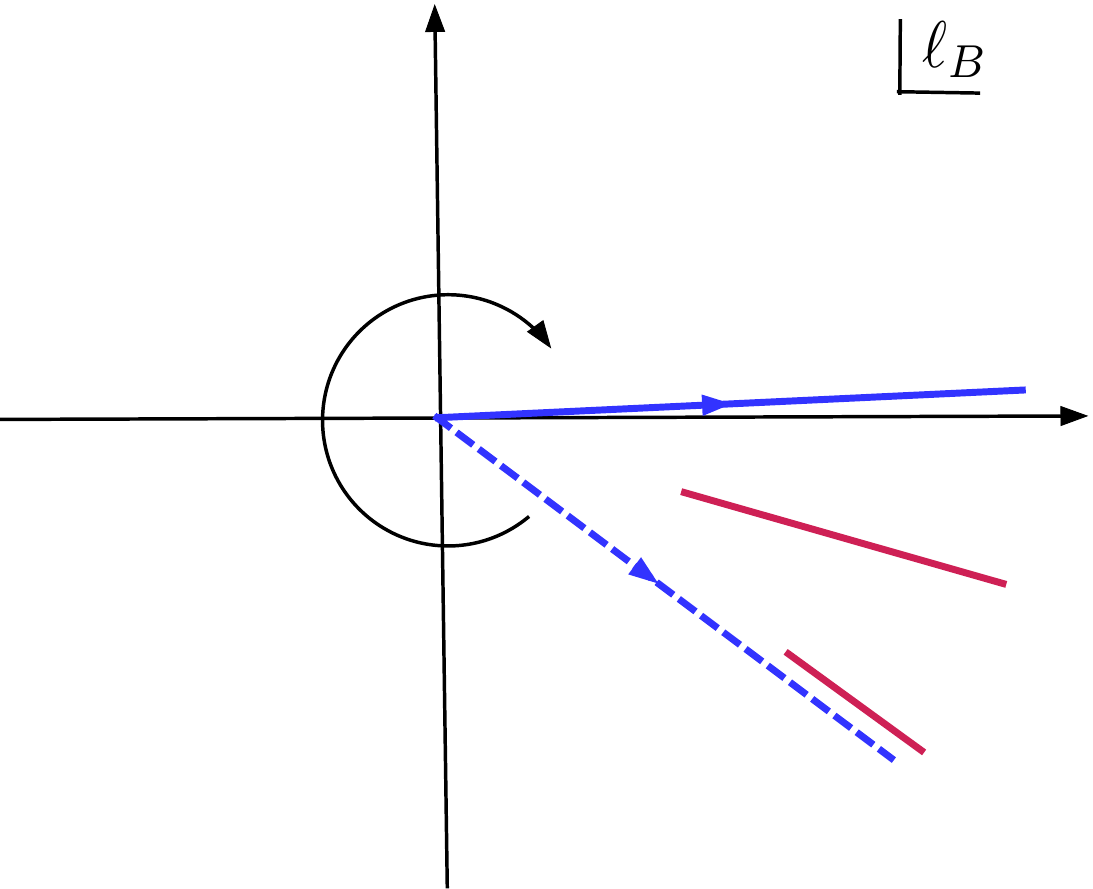}{7}{\small{\textsf{Here we illustrate the argument for rotating the $\ell_B^+$ contour for $\tau_a < \tau_b$. After the rescaling $\ell^+_B \to \ell_B^+e^{-i\tau_b}$, the $\ell_B^+$ contour lies along the dashed blue line. The branch cuts lie along the red lines from $\ell_B^+ = (-2Y_B.Y_a)e^{-i\tau_a}$ to $\infty$ and $\ell_B^+ = (-2Y_B.Y_b)e^{-i\tau_b+i\he}$ to $\infty$. The contour and can therefore be rotated back to the positive real axis without crossing the cuts. A similar argument works for $\tau_a>\tau_b$.}}}

\subsubsection*{Vertical Contours}
In the above calculation, we claimed that the integral $I$ along the contours $C_{\pm}$ vanishes in the limit where the bulk point approaches $\mathcal{H}^{\pm}$; we now show this explicitly. Let us focus on the integral along $C^+$,
\beq\label{SInt0}
I(C_{+}) = i\,\lim_{\ell_B^{-} \to 0 }\int_{s_*^{+}-\varepsilon'}^{s_*^{+}-\varepsilon'+2\pi  i} \frac{ds}{4\sinh^2\left(s/2\right)}\,K_{E}(r_B, is_B ,Y_B | is,Y_b)\,,
\eeq
where as before, $s_*^{+}$ is the time of the future light-like singularity.\footnote{ We have set $\varepsilon, \he \to 0$ as they are unimportant in the present discussion. The new regulator $\varepsilon'$ is required in order to respect causality.} Once again, it is convenient to use the light-cone coordinates $\ell_B^{\pm}$ instead of $(r_B,s_B)$. We then use translation invariance of $K_E$ in Rindler time and transform the integration variable to $w=e^{s-s_*^+}$, so that the above integral becomes
\beq \label{SInt1}
I(C_+) = i \lim_{\ell_B^- \to 0} \oint_{\Gamma} dw\frac{e^{-s_*^+} }{(w-e^{-s^+_*})^2}\,K_E(\frac{\ell^+_{B}}{we^{s_*^+}}, \ell_{B}^- we^{s_*^+}, Y_B | 0,Y_b)
\eeq
where $\Gamma$ is the contour $|w| = 1-\epsilon$. Now in the limit $\ell_B^- \to 0$, we have the following expansions in terms of $\ell_B^-$:
\beq
e^{s^+_*} =\frac{(-2Y_B.Y_b)}{\ell_B^-} + \left(\frac{ \ell_B^+}{2 Y_B\cdot Y_b} -  \ell_B^+\,Y_B\cdot Y_b\right) + \cdots
\eeq
\beq
e^{-s^+_*} =\frac{\ell^-_B}{(-2Y_B.Y_b)} \Big(1 - \ell^+_B\ell_B^-\left(\frac{1}{2} - \frac{1}{(-2 Y_B\cdot Y_b)^2} \right)+\cdots \Big).
\eeq
Using these and expanding out $I(C_+)$, we find
\beq \label{SInt1b}
I(C_+) = i\oint_{\Gamma} dw\left(\frac{e^{-s_*^+} }{w^2}+2\frac{e^{-2s_*^+} }{w^3}+\cdots \right)\,K_E(\frac{\ell_{B}^+}{we^{s_*^+}}, \ell^{-}_Bwe^{s_*^+}, Y_B | 0,Y_b) .
\eeq
In the limit $\ell_B^- \to 0$, it is clear that the above integral effectively probes the bulk field near the Ryu-Takayanagi surface. So it suffices to expand the bulk-to-boundary propagator
\beq
K_{E}(\frac{\ell_B^+}{we^{s_*^+}}, \ell_B^-we^{s_*^+}, Y_B )=K_{E}(0, 0, Y_B )+ \frac{\ell_{B}^+}{we^{s_*^+}}\pa_+K_{E}(0, 0, Y_B )+\ell_{B}^-we^{s_*^+}\pa_-K_{E}(0, 0, Y_B ) + \cdots
\eeq
where we have suppressed the boundary point in the above expressions. Finally we can perform the $w$-integral in equation \eqref{SInt1b} using the residue theorem. The result is
\beq
I(C_+) = -2\pi  \ell_B^- \pa_-K_E(0,0, Y_B| 0, Y_b) + O({\ell_B^-}^2)\,.
\eeq
Therefore, it is clear that $I(C_+)$ vanishes in the limit $\ell_B^- \to 0$. One might worry that the term linear in $\ell_B^-$ contributes inside the symplectic flux $W_{\mathcal{H}^+}$, but it is straightforward to check that the symplectic flux evaluated on $\mathcal{H}^+$ has no derivatives with respect to $\ell_B^-$, and so this does not happen. A similar argument also shows that $I(C_-)$ vanishes on the past horizon. This concludes our proof that the vertical contours do not contribute in the scalar case. However, as we will see in the next section, vertical contours play a crucial role in the case of stress-tensor perturbations, giving rise to boundary terms localized on the Ryu-Takayanagi surface.

\subsection{Relative entropy for stress tensor deformation}\label{sec3.4}

We now want to repeat the above calculation in the case where the excited state is created by turning on a Euclidean source for the stress tensor. Much of the discussion in the previous section carries through directly in this case. Starting from our expression (\ref{re2}), the terms in the relative entropy at second order in the stress-tensor deformation are given by
\beq
\delta^{(2)}S(\rho_A || \rho_A^{(0)})_{grav}  =-\int d\bs\mu_{\a\b\c\d}\, \int_{-\infty}^{\infty} \frac{ds}{4\sinh^2\left(\frac{s+i\he\,\mathrm{sgn}(\tau_a-\tau_b)}{2}\right)}\Big\langle T^{\a\b}(\tau_a,Y_a)  T^{\c\d}(\tau_b+is,Y_b)\Big\rangle
\eeq
where
\small
\beq\label{eq:measure2}
\int d\bs\mu_{\mu\nu\rho\sigma} = \frac{1}{4}\int_0^{2\pi} d\tau_a \int_{\mathbb{H}^{d-1}}dY_a \int_0^{2\pi}d\tau_b \int_{\mathbb{H}^{d-1}}dY_b \,\lambda_{\mu\nu}(\tau_a,Y_a) \,\lambda_{\rho\sigma}(\tau_b,Y_b) \Omega^{-2}(\tau_a,Y_a)\Omega^{-2}(\tau_b,Y_b).
\eeq
\normalsize
Following our discussion in the scalar case, we now rewrite the stress-tensor two-point function in terms of the asymptotic symplectic flux
\beq\label{gf}
-\Big\langle T^{\mu\nu}(\tau,Y_a)  T^{\rho\sigma}(is,Y_b)\Big\rangle = \frac{\widetilde{C}_T}{a^*} \; W^{grav}_{\mathcal{D}(A)} (K_{E;\,mn}^{\mu\nu}, K_{R;\,pq}^{\rho\sigma})
\eeq
where $W^{grav}_{\mathcal{D}(A)}$ is the integral of the symplectic 2-form \eqref{eqn:grav_symplectic_form} associated with the graviton integrated over $\mathcal{D}(A)$, namely the domain of dependence of the boundary region $A$. This is analogous to \eqref{sf1}. Further, $ K^{\a\b}_{E;\,mn}(r_B,is_B, Y_B| \tau,Y_a)$ is the graviton bulk-to-boundary propagator sourced in Euclidean time with the bulk point analytically continued to real time,  and $K_{R;\,pq}^{\rho\sigma}(r_B,s_B, Y_B| s,Y_b)$ is the retarded bulk-to-boundary propagator. Here $\a,\b \cdots$ denote boundary indices and $m, n \cdots$ denote bulk indices; we will often suppress the boundary indices entirely to avoid confusion. We have also extracted a normalization factor $\frac{\widetilde{C}_T}{a^*}$ in \eqref{gf}.
This normalization is due to the following fact. The asymptotic symplectic flux defined through the form $\bs\omega_{grav}$ in \eqref{eqn:grav_symplectic_form} computes a stress-tensor two-point function in an AdS space whose scale is set in terms of $a^*$ as in \eqref{eq:astar}. However, the stress-tensor two-point function in our actual CFT may be normalized independently by a constant conventionally called $C_T$, which is related to $\widetilde{C}_T$ via the first equation in \eqref{eq:ca}.\footnote{ The analogous gravity statement is that the graviton propagator defined by expanding the gravitational Lagrangian to second order in the perturbation, is normalized in a way that is a priori independent of the AdS scale \cite{Haehl:2015rza}. In higher derivative theories of gravity, the ratio $\frac{\widetilde{C}_T}{a^*}$ becomes a function of the higher derivative couplings, see \S\ref{sec:discussion}.} The normalization factor in \eqref{gf} accounts for this and ensures that the right hand side of that equation computes the CFT two-point function for arbitrary values of $\widetilde{C}_T$.

One additional subtlety absent in the scalar case, but which we must confront presently, is that we have to pick a gauge for the graviton bulk-to-boundary propagators. For our purposes, it is important that the propagators have nice analytic properties. A convenient gauge from this point of view is the generalized de-Donder gauge
\beq
\nabla^{(0)}_{m} h^{mn} = 0,\qquad g^{mn}_{(0)}  h_{mn} = 0.
\eeq
In this gauge, we can write the Euclidean bulk-to-boundary propagator conveniently using the index-free notation introduced in \cite{Costa:2014kfa} as
\beq\label{gp}
K_E(X_B, Z_B | P, Z)=  C\frac{\Big(2(Z_B\cdot P)(Z \cdot X_B) - 2 (P\cdot X_B) (Z_B \cdot Z)\Big)^2}{(-2P\cdot X_B)^{d+2}}
\eeq
where $Z_B,Z$ are respectively bulk and boundary auxiliary vectors in embedding space, and
\beq
X_B = \left(r_B Y_B^I , \sqrt{r_B^2-1} \cos\,\tau_B, \sqrt{r_B^2-1} \sin\,\tau_B , r_B Y^m_B\right)
\eeq
is the bulk point written in embedding space, while
\beq
P = \left( Y^I,  \cos\,\tau, \sin\,\tau , Y^m\right)
\eeq
is the boundary point written in embedding space (see Appendix \ref{appA} for further details). Taking derivatives with respect to the auxiliary vectors $Z_B,Z$ and performing appropriate projections restores the four indices of the propagators \cite{Costa:2014kfa}. The propagator \eqref{gp} has nice \emph{causal} properties. In particular, the corresponding Wightman propagators have no acausal brach-cuts; in fact, the only singularities are simple poles at $s_*^{\pm}+2\pi i \mathbb{Z}$. In a different gauge (the holographic gauge for instance), the Wightman propagators can have acausal (spacelike) cuts which could lead to additional complications in our calculation. For instance, recall from our discussion in the scalar case that in performing the $s$-integral we had to resort to a number of contour deformations, which heavily rely on the analytic properties of the propagator. In the presence of acausal cuts, these arguments need further consideration. Fortunately, making the gauge choice \eqref{gp} avoids such issues. Of course, equation \eqref{gf} is gauge-invariant and can be verified in any gauge, but the manipulations we will perform below require the stated gauge-choice.

The relative entropy to second order is then given by
\beq \label{eq:S2fullGrav}
\delta^{(2)}S(\rho_A || \rho_A^{(0)})_{grav}  =\frac{\widetilde{C}_T}{a^*} \int d\bs\mu_{\mu\nu\rho\sigma}\, \int_{-\infty}^{\infty} \frac{ds}{4\sinh^2\left(\frac{s+i\he\,\mathrm{sgn}(\tau_a-\tau_b)}{2}\right)}W^{grav}_{\mathcal{D}(A)} (K_{E;\,mn}^{\mu\nu}, K_{R;\,pq}^{\rho\sigma})
\eeq
Now we can re-trace our steps from the previous section. We push the symplectic flux from the asymptotic boundary to the horizon, and then carry out the $s$-integral using exactly the same techniques used in the scalar case. The integral has three potential contributions (recall figure \ref{fig:fig34.pdf} for a visualization of the contours):
\begin{enumerate}
\item[(i)] from the double-pole coming from $\sinh^2$,
\item[(ii)] from the vertical contour $C_-$ at $s^*_-$ when the bulk point approaches the past horizon,
\item[(iii)] and from the vertical contour $C_+$ at $s^*_+$ when the bulk point approaches the future horizon.
\end{enumerate}
The contribution (i) from the double pole can be computed in essentially the same way as in the scalar case, and one obtains the following analogue of \eqref{eq:scalarFinal}:
\beq \label{CE}
\delta^{(2)}S(\rho_A || \rho_A^{(0)})_{grav}\Big{|}_{pole} = \frac{\widetilde{C}_T}{a^*} \,W^{grav}_{\mathcal{H}^+}\left( h, \mathcal{L}_{\xi_A}h\right) \equiv \frac{\widetilde{C}_T}{a^*} \int_{{\cal H}^+} \bs \omega_{grav}\left(h,{\cal L}_{\xi_A} h\right) \,,
\eeq
where
\beq
h_{mn}(\ell^+_B, Y_B) = \frac{1}{2} \int d\tau dY\, \lambda_{\mu\nu}(\tau, Y) \Omega^{-2}(\tau, Y) K_{E;\,mn}^{\mu\nu}( \ell^+_B, Y_B | \tau, Y) \; .
\eeq
The latter expression is the first-order metric perturbation on our auxiliary AdS space satisfying the linearized Einstein equations and having a Euclidean continuation whose asymptotic behaviour is determined by the sources $\lambda_{\mu \nu}$. Equivalently \cite{MPRV}, it is the source-free Lorentzian solution with asymptotic behavior determined by the CFT stress tensor expectation values via (\ref{holoDict}). This is exactly the metric perturbation $\delta g^{(1)}$ that computes the ball entanglement entropies to first order via HRRT.

The result (\ref{CE}) has the same form as the gravitational part of our desired expression (\ref{CFTresult1}), i.e. the bulk canonical energy in the region $\Sigma_A$.
However, equation \eqref{CE} cannot be whole answer. As reviewed in \S\ref{sec2}, this is the complete expression for canonical energy only in the Hollands-Wald gauge. The calculations in this section make use of a different gauge, and in this case one expects additional boundary terms localized on the Ryu-Takayanagi surface $\widetilde{A}$, which make the full answer gauge-invariant. We will now show that the contributions from the vertical contours $C_{\pm}$ in the CFT calculation correctly reproduce the expected boundary terms, thus restoring gauge-invariance.

\subsubsection*{Vertical contours}
Let us consider the contour integral along $C_+$ for concreteness
\beqn\label{int0}
I_{mn}(C_+) &=& i \lim_{\ell_B^-\to0} \int_{s_*^{+}-\varepsilon'}^{s_*^{+} - \varepsilon'+2\pi i} \frac{ds}{4\sinh^2(s/2)}\, h_{pq}(\ell_B^{+}e^{-s}, \ell_{B}^-e^{s}, Y_B ){J^{p}}_m {J^{q}}_n\nonumber\\
&=& i\lim_{\ell_B^-\to 0} \oint_{\Gamma} dw\frac{e^{-s_*^+} }{(w-e^{-s^+_*})^2}\, h_{pq}(\frac{\ell_{B}^+}{we^{s_*^+}}, \ell_{B}^-we^{s_*^+}, Y_B ) {J^{p}}_m {J^{q}}_n
\eeqn
where as before, $s_*^{+}$ is the time of the future light-like singularity in the bulk-to-boundary propagator, and $\Gamma$ is the contour $|w| = 1-\varepsilon'$. We have also used translation invariance in Rindler time to rewrite this integral directly in terms of the bulk graviton (instead of the bulk-to-boundary propagator). The novel feature in the present case is the appearance of the Jacobian factors on account of the graviton spin
\beq
{J^{\bar\b}}_{\bar\a} = \left(\begin{matrix} \frac{e^{-s_*^+}}{w} & 0 \\ 0 & w e^{s^+_*} \end{matrix}\right),\;\;\; {J^{\b}}_{\a} = \delta^{\b}_{\a},
\eeq
where the unbarred indices $\a, \b$ run along $\mathbb{H}^{d-1}$ while the barred indices $\bar\a, \bar \b $ run over the light-cone directions $\ell_B^{\pm}$. Note that although the bulk point in equation \eqref{int0} is at finite $\ell_B^+$ and small $\ell_B^-$, the boost factors $e^{\pm s^+_*}$ drag the point back to small $\ell_B^+$, while the $w$ integral gets a non-trivial contribution only from the small $\ell_B^-$ region. So effectively the integral $I(C_+) $ probes the geometry only very close to the Ryu-Takayanagi surface.

Let us now evaluate the $w$-integral component by component. Consider for instance $I_{--}$:
\beq \label{int2}
I_{--}(C_+) = i \lim_{\ell_B^- \to 0} \oint_{\Gamma} dw\left(\frac{e^{-s_*^+} }{w^2}+2\frac{e^{-2s_*^+} }{w^3}+\cdots \right)\,w^2 e^{2s^+_*}\, h_{--}(\frac{\ell_B^{+}}{we^{s_*^+}}, \ell_B^{-}we^{s_*^+}, Y_B ) .
\eeq
Next, we expand out $ h_{--}$
\beq
 h_{--}(\frac{\ell_B^{+}}{we^{s_*^+}}, \ell_B^{-}we^{s_*^+}, Y_B )= h_{--}(0, 0, Y_B )+ \frac{\ell_B^{+}}{we^{s_*^+}}\pa_+ h_{--}(0, 0, Y_B )+\ell_B^{-}we^{s_*^+}\pa_- h_{--}(0, 0, Y_B ) + \cdots
\eeq
and then perform the $w$-integral by the residue theorem. This gives
\beq
I_{--}(C_+) = -2\pi  \left[ 2 \left( h_{--}\right)_{\widetilde A} + \ell_B^+ \left(\pa_+ h_{--} \right)_{\widetilde A}+O(\ell_B^-) \right] \,,
\eeq
where the subscript $\widetilde{A}$ stands for evaluation at the undeformed Ryu-Takayangi surface $\ell_{B}^+ = \ell_B^- = 0$. We can also compute the other components similarly, keeping in mind that the Jacobian factors will be different. The final result is
\beq \label{vc1}
I_{--}(C_+) = -  2\pi  \left[ 2 \left( h_{--}\right)_{\widetilde A} + \ell_B^+ \left(\pa_+ h_{--} \right)_{\widetilde A}+O(\ell_B^-)\right]\,,
\eeq
\beq \label{vc2}
I_{-\alpha }(C_+)=  -2\pi \left( h_{-\alpha}\right)_{\widetilde A}+ O(\ell_B^-),
\eeq
\beq \label{vc3}
I_{\a\b}(C_+) =  -2\pi \,\ell_B^- \left(\pa_- h_{\a\b}\right)_{\widetilde{A}}+O((\ell_B^-)^2),
\eeq
\beq \label{vc4}
I_{+-}(C_+) = O(\ell_B^-),
\eeq
\beq \label{vc5}
I_{+\alpha }(C_+) =  O((\ell_B^-)^2),
\eeq
\beq \label{vc6}
I_{++}(C_+)= O((\ell_B^-)^3).
\eeq
In contrast with the scalar calculation, we now see that the integral along the vertical contour $C_+$ has a finite contribution in the limit the bulk point approaches the future horizon. Although we did not display all the $O(\ell_B^-)$ terms above, it is possible to compute them directly from the $w$ contour integral. We have shown only one such term which will be relevant for our purposes, we do not need to compute the rest. A completely analogous calculation determines the integral $I_{mn}(C_-)$.

To summarize, taking into account contributions from all parts of the integration contour, the relative entropy \eqref{eq:S2fullGrav} to second order in a stress tensor deformation is\footnote{ Note that the integrals $I(C_\pm)$ contribute with a minus sign. This is because we previously added them by hand to complete the integration contour in figure \ref{fig:fig34.pdf}. To account for this, we should now subtract these contributions.}
\begin{align} \label{eq:d2Scft}
&\delta^{(2)} S(\rho || \rho_A^{(0)})_{grav} \\
&\qquad= \frac{\widetilde{C}_T}{a^*} \, \bigg\{ \int_{\Sigma_A}\bs{\omega}_{grav} \left( h, \mathcal{L}_{\xi_A}  h\right)
 +\int_{\mathcal{H}^+} \bs{\omega}_{grav} \left( h,- I(C_+) \right) + \int_{\mathcal{H}^-} \bs{\omega}_{grav} \left( h, -I(C_-) \right) \bigg\}\,.   \nonumber
\end{align}

\subsubsection*{Matching the gauge dependent boundary terms}

Now we wish to identify the last two terms in \eqref{eq:d2Scft} with the boundary term localized on the RT surface discussed in \S\ref{sec:gauge_omega} which is the result of the gauge transformation from Hollands-Wald to de-Donder gauge. We recall from \eqref{omega_trans2} that this boundary term had the form
\begin{align} \label{eq:chiRe}
\int_{\widetilde{A}}\, \bs\chi \Big( h ,[\xi_A,V]\Big) = \int_{\Sigma_A} \bs\omega_{grav}(h, \cL_{[\xi_A,V]} g^{(0)} ).
\end{align}
Since this vector field must generate a gauge transformation from a Hollands-Wald gauge, it must obey the Hollands-Wald conditions listed in \S\ref{sec:HW} or in \cite{Lashkari:2015hha}. We will now show that the last two terms in \eqref{eq:d2Scft} can be written in the form \eqref{eq:chiRe}.

In order to proceed, we begin by noting that equations \eqref{vc1} and \eqref{vc2} can also be written as
\beq
I_{--}(C_+) = \left(\mathcal{L}_{\xi_A}  h_{--}\right)_{\widetilde{A}} + 2\pi \frac{\ell_B^ +}{2}\left(\nabla^{(0)}_\a{ h^\a}_{-}-\frac{1}{2} \nabla^{(0)}_{-}{ h^\a}_{\a}\right)_{\widetilde{A}}+O(\ell_B^-)
\eeq
\beq
I_{-\alpha }(C_+)= \left(\mathcal{L}_{\xi_A}  h_{- \alpha}\right)_{\widetilde{A}}+ O(\ell_B^-)
\eeq
where in the former equation we have used the generalized de-Donder gauge conditions. In this form, it is evident that the corresponding contribution to the CFT relative entropy is closely related to the boundary terms on the gravity side. We can make this more precise.
We now claim that there exists a vector field ${\cal V}_{(+)}^a$, such that
\begin{align} \label{eq:chiRe2}
\int_{{\cal H}^+} \bs\omega_{grav}(h, -I(C_+) )  = \int_{{\cal H}^+} \bs\omega_{grav}(h, \cL_{[\xi_A,{\cal V}_{(+)}]} g^{(0)} ) \,.
\end{align}
In practice, we determine this vector field by making an ansatz for ${\cal V}_{(+)}$ and matching $-I_{mn}(C_+)$ with $\cL_{[\xi_A,{\cal V}_{(+)}]} g^{(0)}$. The latter two expressions cannot be matched perfectly, but the remainder term takes a form that drops out of the symplectic flux-integral over ${\cal H}^+$ and is therefore irrelevant. It turns out that the only conditions on ${\cal V}_{(+)}$ we need to require for this matching to work are
\beqn \label{cons1cal}
2\left(\pa_- \cW_{(+)}^+\right)_{\widetilde{A}} &=& \left(\mathcal{L}_{\xi_A} h_{--}\right)_{\widetilde{B}}\\
\left(\frac{1}{2}\pa_{\a} \cW_{(+)}^+ + \frac{\delta_{\a\b}}{u^2}\,\pa_- \cW_{(+)}^{\b}\right)_{\widetilde{A}} &=& \left(\mathcal{L}_{\xi_A} h_{-\a}\right)_{\widetilde{A}}\nonumber\\
\left(\nabla^{(0)}_{\a}\nabla^{(0),\a} \cW_{(+);-} + \left[\nabla^{(0)}_{\a}, \nabla^{(0)}_-\right]\cW_{(+)}^{\a}\right)_{\widetilde{A}}&=& -\left(\nabla^{(0)}_\a{ h^\a}_{-}- \frac{1}{2}\nabla^{(0)}_{-}{ h^\a}_{\a}\right)_{\widetilde{A}}.\nonumber
\eeqn

We recognize these conditions as being precisely part of the Hollands-Wald gauge conditions. We conclude that the components ${\cal V}_{(+)}^+ \big{|}_{\widetilde{\cal A}}$ and $\partial_- {\cal V}_{(+)}^+ \big{|}_{\widetilde{\cal A}}$ coincide with the same components of the Hollands-Wald gauge vector field $V$, see \eqref{cons1}. All other components of ${\cal V}_{(+)}$ are irrelevant for the matching \eqref{eq:chiRe2}.

To summarize, we can make the replacement $-I(C_+) \to  \mathcal{L}_{\left[\xi_A, \cW_{(+)}\right] } g^{(0)}$ inside the symplectic flux on $\mathcal{H}^+$ in our CFT calculation subject to the above constraints on $\cW_{(+)}$. The contribution to the relative entropy at second order from $C_+$ is then given by
\beq
\delta^{(2)}S(\rho_A ||\rho_A^{(0)})_{grav}\Big{|}_{C^+} = \frac{\widetilde{C}_T}{a^*}  \int_{\mathcal{H}^+} \bs{\omega}_{grav} \left(h, \mathcal{L}_{[\xi_A,\cW_{(+)}]}g^{(0)} \right) = \frac{\widetilde{C}_T}{a^*}  \int_{\widetilde{A}} \bs{\chi} \left( h, [\xi_A,\cW_{(+)}] \right)
\eeq
and thus entirely consists of boundary terms localized on the undeformed RT surface.

A similar argument applies to the integral over ${\cal H}^-$.
Therefore, the full CFT result \eqref{eq:d2Scft} is given by
\beq
\begin{split}
& \delta^{(2)} S(\rho || \rho_A^{(0)})_{grav} \\
& = \frac{\widetilde{C}_T}{a^*} \left\{ \int_{\Sigma_A}\bs{\omega}_{grav} \left( h, \mathcal{L}_{\xi_A}  h\right) +\int_{\mathcal{H}^+} \bs{\omega}_{grav} \left( h, \mathcal{L}_{[\xi_A,\cW_{(+)}]}g^{(0)} \right) + \int_{\mathcal{H}^-} \bs{\omega}_{grav} \left( h, \mathcal{L}_{[\xi_A,\cW_{(-)}]}g^{(0)} \right) \right\}\nonumber\\
&  = \frac{\widetilde{C}_T}{a^*} \left\{ \int_{\Sigma_A}\bs{\omega}_{grav} \left( h, \mathcal{L}_{\xi_A}  h\right) +\int_{\widetilde{A}}\Big( \bs{\chi} \left( h, [\xi_A,\cW_{(+)}] \right) + \bs{\chi} \left( h, [\xi_A,\cW_{(-)}] \right)\Big) \right\}\,.
\end{split} \label{cftresult}
\eeq

Defining a vector field $V$ such that
\beq\label{vm1}
\left. V^+\right|_{\widetilde{A}}  = \left. \cW_{(+)}^+\right|_{\widetilde{A}},\;\;\; \left.  \pa_- V^+\right|_{\widetilde{A}} = \left. \pa_- \cW_{(+)}^+\right|_{\widetilde{A}},
\eeq

\beq \label{vm2}
\left.V^-\right|_{\widetilde{A}}=\left. \cW_{(-)}^-\right|_{\widetilde{A}} ,\;\;\; \left. \pa_+V^-\right|_{\widetilde{A}} = \left. \pa_+ \cW_{(-)}^-\right|_{\widetilde{A}},
\eeq
we find that \eqref{cons1cal} and the analogous equations for ${\cal V}_{(-)}$ are precisely the Hollands-Wald constraints that our gauge transformation must satisfy. Since these are the only components of $V$ which enter the boundary term, our CFT result can be written as
\begin{align}\label{eq:d2Sfinal}
\delta^{(2)} S(\rho || \rho_A^{(0)})_{grav} =\frac{\widetilde{C}_T}{a^*} \left\{  \int_{\Sigma_A} \bs{\omega}_{grav} \left( h, \mathcal{L}_{\xi_A}  h\right) + \int_{\widetilde{A}} \bs{\chi} \left( h, [\xi_A,V] \right) \right\}\,,
\end{align}
where $V$ generates a gauge transformation from a Hollands-Wald gauge to the de-Donder gauge.

\subsection{Summary of the CFT result}

Combining the results from \S\ref{sec3.2} and \S\ref{sec3.4} for a scalar and a stress tensor deformation of the Euclidean path integral, we find:
\begin{align} \label{summary}
\delta^{(2)}S(\rho_A||\rho_A^{(0)}) =  \int_{\Sigma_A}   \Big\{\frac{\widetilde{C}_T}{a^*} \,\bs\omega_{grav} \big( h ,\cL_{\xi_A} h \big) + \bs\omega_{\phi}\left( \delta\phi,  \mathcal{L}_{\xi_A} \delta\phi\right) \Big\}
+ \frac{\widetilde{C}_T}{a^*} \int_{\widetilde{A}} \bs\chi \big( h , [\xi_A,V] \big) \, ,
\end{align}
where $\delta\phi$ is the expectation value of the bulk scalar field dual to the operator sources in the Euclidean path integral and similarly $h$ is the expectation value of the perturbation to the bulk metric in Hollands-Wald gauge.  While we have performed the above CFT calculation for a particular ball-shaped region, conformal symmetry dictates that the same calculation goes through more generally for any ball-shaped region of arbitrary radius, or in an arbitrary boosted frame, with appropriately chosen conformal factors.

Now, recall the fundamental gravitational identity \eqref{eqn:grav_result2},
\begin{align}
&(E^{grav}_A - S^{grav}_A)^{(2)} \label{eqn:grav_result2rep}\\
&\qquad =
\int_{\Sigma_A} \Bigg\{ \bs\omega_{grav} \Big( h ,\cL_{\xi_A}h  \Big) + \bs\omega_{\phi} \Big( \delta\phi ,\cL_{\xi_A} \delta  \phi\Big)
- 2 \xi_A^a \left( E_{ab}^{(2)}\right) \bs\epsilon^b
 \Bigg\}+ \int_{\widetilde{A}} \bs\chi \Big( h , [\xi_A,V] \Big) , \nonumber
\end{align}
Putting together the above equations (asserting \eqref{eq:ca}, i.e., $\frac{\widetilde{C}_T}{a^*}=1$) and using the basic ingredient that $\delta^{(2)}S(\rho_A||\rho_A^{(0)}) = (E^{grav}_A - S^{grav}_A)^{(2)} $ we find
\beq
 \int_{\Sigma_A}  \xi_A^a \, \bs\epsilon^b \; E_{ab}^{(2)} = 0 \,.
\eeq
From the arbitrariness of the choice of slice $\Sigma_A$, one infers by the argument in \cite{Faulkner:2013ica} (or alternatively \cite{Mosk:2016elb,Czech:2016tqr}) the vanishing of the integrand, i.e., the second order Einstein equations (including matter sources) as anticipated in \eqref{eq:EFE}. This establishes the central result of our work.

Without assuming $\frac{\widetilde{C}_T}{a^*}=1$, the above argument obviously fails. In the next section we shall briefly discuss how the derivation needs to be modified in this case.

\section{Discussion}
\label{sec:discussion}

In this paper, we have seen the emergence of non-linear gravitational dynamics from the physics of entanglement entropy in conformal field theories. For a wide class of states designed to give coherent states of bulk fields in the case of holographic theories, we have seen that $(i)$ there exists an asymptotically AdS spacetime, $M(\pert)$, which correctly computes via HRRT ball entanglement entropies for the state (and also one-point functions and relative entropy compared with the vacuum), and $(ii)$ any such geometry satisfies Einstein's equations sourced by matter fields determined in terms of the CFT one-point functions. Both results hold up to second order in the sources used to define the state; in particular, the gravitational equations that emerge hold to second order in perturbations about AdS and include nonlinear terms governing the self-interaction of the metric and the interaction of the metric with bulk matter fields.

Our result is universal for theories with $\widetilde{C}_T = a^*$ (this covers all theories in $d=2$) and does not rely on any assumptions about holography. In particular, both the existence of a geometry capturing the entanglement entropies and the fact that such a geometry satisfies local bulk equations are outputs of our calculations. On the other hand, as we now explain, our result provides a consistency check for the AdS/CFT correspondence and the HRRT formula.

\subsubsection*{Necessity of our results for the AdS/CFT correspondence}

We have seen that for states of the form (\ref{PIstate0}) in any CFT, we write the entanglement entropy to second order in the one-point functions\footnote{ In our calculation, we expressed the second order terms in terms of the first-order sources. However, the sources at first order are linearly related to the one-point functions at first order via a map that depends only on the CFT two-point function. We have $\lambda_{T} \sim \langle T T \rangle \sim (1/\widetilde{C}_T) \langle T\rangle$, leading to the factor of $1/\widetilde{C}_T$ here. If we instead express the result directly in terms of sources, the first order term $\left< T(x) \right>$ when expanded at second order in the sources is sensitive to the full CFT stress tensor 3 point function which depends on more parameters in addition to $\widetilde{C}_T$ and $a^\star$. However this dependence cancels from the entropy when expressed in terms of  one-point functions.}
\begin{equation}
\begin{split}
\label{SCFT}
S_{CFT}(A,\langle {\cal O} \rangle,\langle T \rangle) = a^* S^{(0)}(A)  + \int K^{(1)}_A(x) & \langle T(x) \rangle  + \int \int K^{(2)}_A (x_1,x_2) \langle O_\alpha(x_1) \rangle \langle O_\alpha(x_2) \rangle \\ &\quad\;\,+  {1 \over \widetilde{C}_T} \int \int L^{(2)}_A(x_1,x_2) \langle T(x_1) \rangle \langle T(x_2) \rangle + \dots \;
\end{split}
\end{equation}
up to second order, this expression depends on which CFT we are dealing with only through the coefficients $a^*$ and $\widetilde{C}_T$.

Starting from the same state, we can perturbatively define an asymptotically AdS geometry ${\cal M}_{\langle {\cal O} \rangle}(\pert)$ to second order by using (\ref{eq:astar}) to define the AdS scale and (\ref{holoDict}) and  (\ref{phi_asympt}) to define the asymptotic behavior of the bulk fields. We then solve Einstein's equations (and linearized matter equations) perturbatively to determine the metric in the bulk up to second order. From this, we can apply the HRRT formula to this geometry to define a functional $S_{grav}(A,\langle {\cal O} \rangle,\langle T \rangle)$; this depends only on the parameter $a^*$ and takes the form\footnote{ This gravitational result was previously discussed in \cite{Beach:2016ocq}.}
\begin{equation}
\begin{split}
\label{Sgrav}
S_{grav}(A,\langle {\cal O} \rangle, \langle T \rangle) = a^* \hat{S}^{(0)}(A)  + \int \hat{K}^{(1)}_A(x) &\langle T(x) \rangle  + \int \int \hat{K}^{(2)}_A (x_1,x_2) \langle O_\alpha(x_1) \rangle \langle O_\alpha(x_2) \rangle \\ &\quad\;\;\; +  {1 \over a^*} \int \int \hat{L}^{(2)}_A(x_1,x_2) \langle T(x_1) \rangle \langle T(x_2) \rangle + \dots
\end{split}
\end{equation}

Now, according to the AdS/CFT correspondence, there exist families of CFTs with $a^* = \widetilde{C}_T$ dual to gravitational theories whose classical description (associated to the leading order in the $1/a^*$ expansion of the CFT for large $a^*$) is Einstein gravity coupled to matter. Then for states of the form (\ref{PIstate0}) in such a theory, $S_{grav}(A,\langle {\cal O} \rangle,\langle T\rangle)$ and $S_{CFT}(A,\langle {\cal O} \rangle,\langle T\rangle)$ must agree with each other in the limit of large $a^*$, with vanishing fractional error in the large $a^*$ limit. This is only possible if all the functionals appearing in (\ref{SCFT}) and (\ref{Sgrav}) are the same; it follows immediately that the two functionals also agree for any other theory with $a^* = \widetilde{C}_T$. This is what we have shown directly in this paper; thus, our results provide a consistency check for AdS/CFT.

\subsubsection*{Generalization to theories with $\widetilde{C}_T \neq a^*$}

By the AdS/CFT correspondence, we expect that there also exist families of CFTs dual to gravitational theories whose classical descriptions include various higher curvature terms. For these theories, the CFT coefficients $a^*$ and $\widetilde{C}_T$ are determined by the parameters in the gravitational Lagrangian; in general, we will have $\widetilde{C}_T \neq a^*$. Following the logic in the previous section, if the AdS/CFT correspondence is also correct for these more general theories, the analogue of (\ref{Sgrav}) must again match with (\ref{SCFT}) in the limit of large central charges; in this case $\widetilde{C}_T$ can differ from $a^*$, so we can potentially reproduce the result (\ref{SCFT}) for any $a^*$ and $\widetilde{C}_T$, i.e. the general result for {\it any} CFT, by using a two-parameter family of gravitational Lagrangians.

In practice, we expect that it is not necessary to use gravitational Lagrangians which arise as the dual of some CFT. We can follow the procedure described above to define $S^{grav}$ for any Lagrangian ${\cal L}$ (making use of an appropriately modified holographic dictionary and entanglement functional; see below), and we expect that the result to this order should take the CFT form.\footnote{ If not, we could conclude that this gravity Lagrangian lies in the swampland, i.e. that it does not correspond to the classical limit of a consistent quantum theory.} Thus, we can work with a simple two-parameter theory such as the Gauss-Bonnet theory ${\cal L} = \frac{1}{16\pi G_N} \left[ (R-2\Lambda) - \frac{\alpha_{_{GB}}}{2\Lambda}\frac{d(d-1)}{(d-2)(d-3)} (R_{abcd}R^{abcd} - 4\, R_{ab} R^{ab} + R^2) \right]$, which in $d\geq4$ which leads to \cite{Myers:2010jv}
\beq \label{eq:caGB}
  \frac{\widetilde{C}_T}{a^*} =  1 + \frac{4}{d-3}\, \alpha_{_{GB}} + {\cal O}(\alpha_{_{GB}}^2) \,.
\eeq

We emphasize that in the more general case, to define (\ref{Sgrav}) for some gravitational Lagrangian ${\cal L}$, we should use the correct entanglement entropy functional associated with ${\cal L}$.\footnote{ The relative normalization on the two sides of the holographic dictionary (\ref{holoDict}) also depends on ${\cal L}$ \cite{Faulkner:2013ica}.} It is known that in the presence of higher derivative corrections the functional which computes entanglement entropy needs to be corrected by extrinsic curvature terms \cite{Jacobson:1993vj,Fursaev:2013fta,Dong:2013qoa,Camps:2013zua}. These correction terms are quadratic in extrinsic curvatures and therefore contribute at second order in perturbation theory. We expect that these terms need to be taken into account to correctly match with $S_{CFT}$; it is likely that we can constrain these terms independent from other methods by requiring that this matching works correctly.

In summary, we can make the following conjectures:
{\it
\begin{enumerate}
\item
For any CFT, for any state of the form (\ref{PIstate0}), there exists a gravitational Lagrangian ${\cal L}$ in some two-parameter family and a geometry $M_{\cal L}(\pert)$ defined by a metric $g^{(0)}_{AdS} + \pert\, \delta g^{(1)} + \pert^2 \,\delta g^{(2)} + \dots$ and fields $\phi_\alpha$ associated with the sourced operators, such that the entanglement entropies for all ball-shaped regions are correctly computed from $M_{\cal L}(\pert)$ up to order $\pert^2$ via the generalized HRRT formula with entanglement entropy functional obtained from ${\cal L}$, and the one-point functions are correctly computed to order $\pert^2$ from the asymptotic behavior of the fields. For example,  for any value of $\frac{\widetilde{C}_T}{a^*}$ in $d \geq 4$, one can choose ${\cal L}$ to be the Gauss-Bonnet Lagrangian for a particular value of the Gauss-Bonnet coupling and the cosmological constant.
\item
Any geometry $M_{\cal L}(\pert)$ that correctly computes entanglement entropies for a CFT state of the form (\ref{PIstate0}) using the entanglement entropy formula appropriate to a Lagrangian ${\cal L}$ must satisfy the equation of motion associated with ${\cal L}$, up to second order in $\pert$, where the stress-energy tensor is that for a set of matter fields corresponding to the sourced operators ${\cal O}_\alpha$, and these matter fields solve linearized equations about AdS with boundary conditions specified by the CFT one-point functions of ${\cal O}_\alpha$.
\end{enumerate}
}

To complete the proof of these assertions, we need to show (as we did in section 3 for Einstein gravity coupled to matter) that the CFT relative entropy for ball-shaped regions can also be written in terms of the symplectic form associated with a more general Lagrangian ${\cal L}$. As we explained in section 2, the geometrical identity \eqref{Key2} that we needed also generalizes to arbitrary ${\cal L}$. Combining the CFT calculation with the gravitational identity as before should hopefully lead to the conjectured assertion; however, there are a number of subtleties to sort out in the more general case. We are planning to report on this issue in the future.

\subsubsection*{Higher orders in perturbation theory}

It is interesting to consider extending our results to higher orders in state perturbation. At each order in perturbation theory, we can ask what conditions on the CFT guarantee that the ball entanglement entropies for states of the form (\ref{PIstate0}) can be reproduced via HRRT (or a more general functional) by a spacetime $M(\pert)$ defined perturbatively to the same order, and we can try to extend our derivation to show that $M(\pert)$ must satisfy Einstein's equations (or some more general gravitational equations) at higher orders. We note that the key identity (\ref{Key}) that formed the basis of our proof is already valid non-perturbatively; we have so far exploited it only at first and second order.\footnote{ Ideally, we would like to show that with some set of conditions on the CFT, the full result for the derivative of relative entropy on the left side of (\ref{Key}) matches with the first term on the right side of (\ref{Key}), possibly plus some term that vanishes on shell. This may be enough to imply that the full nonlinear gravitational equations are satisfied everywhere in the geometry reached by HRRT surfaces.} Also, the CFT calculations of relative entropy in terms of the sources can also be extended to higher order; at $n$th order in the sources the expression for relative entropy will involve vacuum $n$ point functions of operators in the CFT.\footnote{ See \cite{Sarosi:2017rsq} for related discussion.} We thus expect that agreement at order $n$ between $S_{CFT}$ and $S_{grav}$ will place constraints on the structure of CFT $n$-point functions for a holographic theory.

For $n=3$, the functional form of the correlators is still fixed by conformal invariance up to a small number of parameters; these should be determined in terms of $a^*$ for agreement with Einstein gravity.\footnote{ For example, the stress-tensor three-point functions are fixed by conformal symmetry up to three parameters (including $C_T$). For the entanglement entropies to be reproduced by a geometry using HRRT, we expect these parameters to be fixed in terms of $C_T$. More generally, this means that if it is possible to reproduce the entanglement entropy for an arbitrary CFT at this order using some geometry, we must consider the entanglement functional associated with at least a three-parameter set of higher derivative couplings.} At $n=4$ and higher, where the functional form of correlation functions can include general functions of cross-ratios, these constraints should be severe. It would be interesting to understand the relation between these constraints obtained this way and other conditions such as the sparseness of the spectrum of light operators.

\acknowledgments
The authors gratefully acknowledge initial collaboration and insightful discussions with Rob Myers. We further thank V. Balasubramanian, T. Hartman, V.\ Hubeny, N.\ Lashkari, A.\ Lewkowycz, M. Mezei, M.\ Rangamani, A. Speranza and B. Stoica for useful comments and discussions. The work of FH, EH, and MVR is supported in part by the Natural Sciences and Engineering Research Council of Canada, a Simons Investigator award from the Simons Foundation, and the ``It From Qubit'' collaboration grant from the Simons Foundation.

FH wishes to thank UC Davis, Simons Center for Geometry and Physics and Stanford University for hospitality and support during the course of this project.

OP and CR wish to acknowledge support from the Simons Foundation (\#385592, Vijay Balasubramanian) through the It From Qubit Simons Collaboration. CR was also supported by the Belgian Federal Science Policy Office through the Interuniversity Attraction Pole P7/37, by FWO-Vlaanderen through projects G020714N and G044016N, and by Vrije Universiteit Brussel through the Strategic Research Program ``High-Energy Physics''.

TF acknowledges support from DARPA grant D15AP00108.

This research was supported in part by Perimeter Institute for Theoretical Physics. Research at Perimeter Institute is supported by the Government of Canada through the Department of Innovation, Science and Economic Development and by the Province of Ontario through the Ministry of Research and Innovation.

\appendix

\section{Coordinates in embedding space and conventions}\label{appA}

We use the following parametrization for AdS$_{d+1}$ embedded in $\mathbb{R}^{1,d+1}$:
\beq
\begin{split}
\text{Poincar\'e coordinates:} \qquad X^A &= \left(\frac{1+z^2+x^2}{2z}, \frac{1- x^2-z^2}{2z},\frac{x}{z}\right) \,, \\
\text{Hyperbolic coordinates:} \qquad X^A &= \left(rY^I, \sqrt{r^2-1} \cos(\tau), \sqrt{r^2-1}\sin(\tau), rY^m\right) \,,
\end{split}
\eeq
where the hyperbolic space $ \mathbb{H}^{d-1}$ is described by:
\beq
 Y \equiv (Y^I,Y^m) = \left(\frac{1+u^2+\vec{x}^2}{2u}, \frac{1- \vec{x}^2-u^2}{2u},\frac{\vec{x}}{u}\right) \in \mathbb{H}^{d-1} \,,
\eeq
such that $Y^I > 0$ and $Y^2 \equiv -(Y^I)^2 + Y^m Y^m = -1$.
Similarly, points on the boundary $S^1\times \mathbb{H}^{d-1}$ are denoted as
\beq
 P \equiv (P^I,P^{II},P^\mu)  = \left( Y^I, \cos(\tau),\sin (\tau), Y^m\right) \in S^1\times \mathbb{H}^{d-1}\,,
\eeq
such that they describe a section of the light-cone $P\cdot P = 0$, $P^I >0$ in the flat embedding space geometry $\eta_{AB} = \text{diag}(-1,1,\ldots,1)$.
This hyperbolic description of the boundary geometry is related to the flat space description
\beq
P = \left(\frac{R^2+x^2}{2R}, \frac{R^2-x^2}{2R}, x^{\mu}\right) \in \mathbb{R}^d
\eeq
by a conformal transformation. The conformal factor associated with this map is
\beq
\Omega(\tau, Y) = R^{-1} (Y^I + \cos \tau).
\eeq

The induced $AdS$ metric (in the hyperbolic black hole coordinates $x^m = (s=-i\tau,r,u,\vec{x})$) is given by
\beq
g^{(0)} = \eta_{MN}\frac{\pa X^M}{\pa x^m}\frac{\pa X^N}{\pa x^n}dx^m dx^n = -(r^2-1)ds^2+\frac{dr^2}{(r^2-1)}+\frac{r^2}{u^2}\left(du^2+ d\vec{x}^2\right) \,,
\eeq
and the Killing field $\xi_A$ in these coordinates is given by $\xi_A = 2\pi \pa_{s}$.

For our calculations, it is much more useful to work in the $\ell_{\pm}$ coordinates defined as
\beq
\ell_{\pm} = \sqrt{r^2-1} \; e^{\pm s}\,.
\eeq
In this case, the Jacobians are given by
\beqn
\frac{\pa X^A}{\pa u} &=& \sqrt{1+\ell_+\ell_-}\left(E_0 - \frac{Y}{u}\right)^A\nonumber\\
\frac{\pa X^A}{\pa \vec{x}^i} &=& \sqrt{1+\ell_+\ell_-}\left(\frac{x^i}{u} E^A_0 + \frac{\delta^A_i}{u}\right)\nonumber\\
\frac{\pa X^A}{\pa \ell_{\pm}} &=& \frac{\ell_{\mp}}{2\sqrt{1+\ell_+\ell_-}}Y^A + \frac{1}{2}E^A_{\pm}
\eeqn
The AdS metric is given by
\beq
g^{(0)} = \frac{1}{4(1+\ell_+\ell_-)}\left(-\ell_-^2d\ell_+^2 - \ell_+^2d\ell_-^2\right)+\left( \frac{1}{4}+\frac{1}{4(1+\ell_+\ell_-)}\right)2 d\ell_+d\ell_- + \frac{1+\ell_+\ell_-}{u^2}\left(du^2+ d\vec{x}^2\right)
\eeq
\beq
g_{(0)}^{-1} = \ell_+^2\pa_+^2 +\ell_-^2\pa_-^2+ (2+\ell_+\ell_-)2\pa_+\pa_- + \frac{u^2}{1+\ell_+\ell_-}\left(\pa_u^2+\pa_{\vec{x}}^2\right)
\eeq
while the Killing field $\xi_A$ in these coordinates is given by $\xi_A =2\pi (\ell_+\pa_{+} - \ell_- \pa_-)$.

To summarize our index conventions: we use capital indices $A,B,\ldots$ for embedding space $\mathbb{R}^{1,d-1}$. Small letters $a,b,\ldots,m,n,\ldots$ are used for AdS$_{d+1}$. Greek indices $\alpha,\beta,\ldots$ are indices along the $(d-1)$-dimensional HRRT surface, and barred indices $\bar\alpha,\bar\beta,\ldots$ denote normal directions $(\ell_B^+, \ell_B^-)$ normal to that surface. We also use indices from the second half of the Greek alphabet $\mu,\nu,\ldots$ for CFT$_d$ boundary coordinates.

\section{Relative entropy at second order}\label{app0}

In this appendix we review the connection between relative entropy and the Fisher information metric and provide some related calculational details skipped in the main text.

\subsubsection*{Fisher information metric}

The relative entropy at second order is given by
\be \label{eq:Fisher}
{1 \over 2} {d^2 \over d \pert^2} S(\rho(\pert)  || \rho(0)) = {1 \over 2} \tr\left(\delta \rho {d \over d \pert} \ln(\rho_0 + \pert \delta \rho)\right) \,,
\ee
where
\be
\rho(\pert) = \rho_0 + \pert \,\delta \rho + \dots \; ,
\ee
and we have used the fact that $\tr(\delta \rho) = 0$. The expression \eqref{eq:Fisher} evaluated at $\pert=0$ defines the Fisher information metric $F(\delta \rho, \delta \rho)$. Using
\be
-\ln(X) = \int_0^\infty {d s \over s} \left(e^{-sX} - e^{-s} \right)  \; ,
\ee
and
\be
{d \over d \pert} e^{A + \pert B} = \int_0^1 dx \,e^{Ax} B e^{(1-x) A} \; ,
\ee
we get
\be
{1 \over 2} {d^2 \over d \pert^2} S(\rho(\pert)  || \rho_0) = {1 \over 2} \int_0^1 dx \int_0^\infty ds\, \tr\left(\delta \rho e^{-s x \rho_0} \delta \rho e^{-(1-x) s \rho_0}\right) \; .
\ee
In a basis where $\rho_0$ is diagonal with eigenvalues $\rho_a$, this gives
\be
\label{REdiag}
{1 \over 2} {d^2 \over d \pert^2} S(\rho(\pert)  || \rho_0) = \sum_{\rho_a < \rho_b} |\delta \rho_{ab}|^2 {\ln(\rho_b) - \ln(\rho_a) \over \rho_b - \rho_a} + {1 \over 2} \sum_{\rho_a = \rho_b} |\delta \rho_{ab}|^2 (\rho_a)^{-1}
\ee
Using contour integration, it is straightforward to show that
\be
{1 \over 4} \int_{-\infty}^{\infty} ds  {e^{i s x \over 2 \pi} \over 1 + \cosh(s)} = {x \over 2 (e^{x \over 2} - e^{-{x \over 2}})}
\ee
Using this, we can show that the following formulae reproduce the results (\ref{REdiag}) above, again by going to a basis where $\rho_0$ is diagonal
\be
\label{FI1}
{1 \over 2} {d^2 \over d \pert^2} S(\rho(\pert)  || \rho_0)  = {1 \over 4} \int_{-\infty}^{\infty} {ds \over 1 + \cosh(s)} \tr (\delta \rho \rho_0^{-{1 \over 2} - {i s \over 2 \pi}} \delta \rho \rho_0^{-{1 \over 2} + {i s \over 2 \pi}}) \,.
\ee
Shifting the integration contour by $s \to s \pm i \pi (1-\epsilon)$, or directly using the results
\be
\int_{-\infty}^\infty ds {e^{i s x \over 2 \pi} \over 4 \sinh^2\left(\frac{s \pm i \epsilon}{2}\right)} = {\pm x \over 1 - e^{\pm x}}
\ee
we obtain the alternative formulae
\be
\label{FI2}
{1 \over 2} {d^2 \over d \pert^2} S(\rho(\pert)  || \rho_0)  = -{1 \over 2} \int_{-\infty}^{\infty} {ds \over 4 \sinh^2\left(\frac{s + i \epsilon}{2}\right)} \tr(\rho_0^{-1} \delta \rho \rho_0^{i s \over 2 \pi} \delta \rho \rho_0^{-i s \over 2 \pi})
\ee
and
\be
\label{FI3}
{1 \over 2} {d^2 \over d \pert^2} S(\rho(\pert)  || \rho_0)  = -{1 \over 2} \int_{-\infty}^{\infty} {ds \over 4 \sinh^2\left(\frac{s - i \epsilon}{2}\right)} \tr(\rho_0^{-1} \delta \rho \rho_0^{-i s \over 2 \pi} \delta \rho \rho_0^{i s \over 2 \pi}) \; .
\ee
We will make use of these results in the following paragraph to express the second order relative entropy in terms of an analytically continued two-point function. As an aside, we note that replacing the two instances of $\delta \rho$ in (\ref{FI1}), (\ref{FI2}), or (\ref{FI3}) with $\delta \rho_1$ and $\delta \rho_2$ (in either order) gives a symmetric bilinear form that defines the Fisher information metric about $\rho_0$.

\subsubsection*{Derivation of Eq.\ \eqref{eq:fisherRes}}

We can use the representations \eqref{FI2} and \eqref{FI3} to derive the expression \eqref{eq:fisherRes} for second order relative entropy in terms of a time ordered two-point function.
In general, we have for state perturbations of the form \eqref{deltarho} the following expression for relative entropy:
\beq\label{eq:d2Stemp}
\begin{split}
& \delta^{(2)}S(\rho_A || \rho_A^{(0)}) = 2\int d^d x_a \int d^d x_b \lambda_\alpha(x_a) \lambda_\alpha(x_b) F(\rho_A^{(0)} {\cal O}_\alpha(x_a), \rho_A^{(0)} {\cal O}_\alpha(x_b)) \,.
\end{split}
\eeq
In anticipation of the time ordering, let us distinguish the two cases $\tau_a>\tau_b$ and $\tau_a < \tau_b$ (where $\tau_{a,b}$ are the Euclidean times of the insertion points of the two scalar operators). For $\tau_a > \tau_b$, we use the representation \eqref{FI2} to write
\beq
\begin{split}
& F(\rho_A^{(0)} {\cal O}_\alpha(x_a), \rho_A^{(0)} {\cal O}_\alpha(x_b))_{\tau_a > \tau_b}\\
&\qquad
 =  -\frac{1}{2}\int_{-\infty}^{\infty} {ds \over 4 \sinh^2 \left({s + i \epsilon \over 2}\right)} \tr \left( {\cal O}_\alpha(x_b)
 e^{-{i s \over 2 \pi} H_A} \, e^{-H_A} {\cal O}_\alpha(x_a) e^{{i s \over 2 \pi} H_A} \right) \\
 &\qquad=  -\frac{1}{2}\int_{-\infty}^{\infty} {ds \over 4 \sinh^2 \left({s + i \epsilon \over 2}\right)} \frac{\Omega^{\Delta}(\tau_b,\tilde{x}_b)}{ \Omega^{\Delta}(\tau_b+is,\tilde{x}_b)} \tr \left( e^{-H_A} {\cal O}_\alpha(\tau_a,\tilde{x}_a) \, {\cal O}_\alpha(\tau_b+is,\tilde{x}_b)  \right) \cr
\end{split}
\eeq
where $\rho_A^{(0)} = e^{-H_A}$ is the vacuum state, which is computed naturally using the Euclidean path integral.
Similarly, when $\tau_a < \tau_b$, we use \eqref{FI3} and get
\beq
\begin{split}
 &F(\rho_A^{(0)} {\cal O}_\alpha(x_a), \rho_A^{(0)} {\cal O}_\alpha(x_b))_{\tau_a < \tau_b}\\
&\qquad =  -\frac{1}{2}\int_{-\infty}^{\infty} {ds \over 4 \sinh^2 \left({s - i \epsilon \over 2}\right)} \frac{\Omega^{\Delta}(\tau_b,\tilde{x}_b)}{ \Omega^{\Delta}(\tau_b+is,\tilde{x}_b)} \tr \left( e^{-H_A} {\cal O}_\alpha(\tau_b+is,\tilde{x}_b) \, {\cal O}_\alpha(\tau_a,\tilde{x}_a)  \right) \cr
\end{split}
\eeq
The sum of these two gives
\beq
\begin{split}
& F(\rho_A^{(0)} {\cal O}_\alpha(x_a), \rho_A^{(0)} {\cal O}_\alpha(x_b))\\
 &\qquad =-\frac{1}{2}  \int_{-\infty}^{\infty} {ds \over 4 \sinh^2 \left({s + i \epsilon {\rm sgn}(\tau_a - \tau_b) \over 2 }\right)} \frac{\Omega^{\Delta}(\tau_b,\tilde{x}_b)}{ \Omega^{\Delta}(\tau_b+is,\tilde{x}_b)} \langle \mathcal{T} [{\cal O}_\alpha(\tau_a,\tilde{x}_a) {\cal O}_\alpha(\tau_b + i s,\tilde{x}_b)] \rangle \,,
\end{split}
\eeq
where $\mathcal{T}$ indicated $\tau$ ordering. Plugging this into \eqref{eq:d2Stemp} gives the result \eqref{eq:fisherRes}.

\section{Derivation of Eq.\ \eqref{eq:chivanish}}
\label{sec:App3}

Here, we give some details on the proof that  $\bs\chi \big(\cL_{\xi_A} ( h + \cL_{V} g ) , V \big)_{\widetilde{A}} = 0$. We abbreviate $\dot{\gamma} \equiv \cL_{\xi_A} ( h + \cL_{V} g)$. The perturbation $h + \cL_{V} g$ is in Hollands-Wald gauge by construction, so $\dot{\gamma}|_{\widetilde{A}}=0$. Using this fact, we can drop terms proportional to $\dot{\gamma}$ and evaluate the desired form:
\beq
\begin{split}
 \bs\chi\big( \dot{\gamma}, \, V \big)_{\widetilde{A}} &= \bs \epsilon_{ab} \big( \nabla^b \dot{\gamma}^a_c \, V^c - \nabla_c \dot{\gamma}^{ac} \, V^b + \nabla^a \dot{\gamma}^c_c \, V^b \big) \\
  &=  2\,\bs \epsilon_{+-} \left( \nabla^{[-} \dot{\gamma}^{+]}_{\bar \alpha}\, V^{\bar \alpha}  - \nabla_c \dot{\gamma}^{c[+} \, V^{-]} + \nabla^{[+} \dot{\gamma}^c_c \, V^{-]} \right) \,,
\end{split}
\eeq
where we used the fact that the only non-vanishing components of the vector $V$ evaluated on the RT surface are $V^+$ and $V^-$. We now lower indices and calculate:
\beq
\begin{split}
  \bs\chi\big( \dot{\gamma}, \, V \big)_{\widetilde{A}}
    &=  -8\,\bs \epsilon_{+-} \left( \nabla_{[-} \dot{\gamma}_{+]\bar{\alpha}}\, V^{\bar{\alpha}}   - \nabla_c \dot{\gamma}^{c}_{[+} \, g_{-]\bar{\alpha}}\,V^{\bar{\alpha}} + \nabla_{[+} \dot{\gamma}^c_c \, g_{-]\bar{\alpha}}\, V^{\bar{\alpha}} \right) \\
    &=  -8\,\bs \epsilon_{+-} \Big\{ \nabla_{[-} \dot{\gamma}_{+]\bar{\alpha}}\, V^{\bar{\alpha}}  - \left(2\nabla_+ \dot{\gamma}_{-[+} + 2\nabla_- \dot{\gamma}_{+[+} + \nabla_\alpha \dot{\gamma}^{\alpha}_{[+} \right) g_{-]\bar{\alpha}}\,V^{\bar{\alpha}}  \\
    &\qquad\qquad\quad  + \left(4\nabla_{[+} \dot{\gamma}_{+-} + \nabla_{[+} \dot{\gamma}^\alpha_\alpha \right)\, g_{-]\bar{\alpha}}\, V^{\bar{\alpha}} \Big\} \\
    &= -8\,\bs \epsilon_{+-} \Big\{ \nabla_{[-} \dot{\gamma}_{+]\bar\alpha} -2 \left(  \nabla_+ \dot{\gamma}_{-[+} +  \nabla_- \dot{\gamma}_{+[+} \right) g_{-]\bar\alpha} +  4\nabla_{[+} \dot{\gamma}_{+-} \, g_{-]\bar\alpha}    \Big\} V^{\bar\alpha} \\
    &= 0 \,,
\end{split}
\eeq
where the last step can easily be checked and in the penultimate step we used
\beq
  \nabla_\alpha \dot{\gamma}^\alpha_\pm \Big{|}_{\widetilde{A}} = \nabla_\pm \dot{\gamma}^\alpha_\alpha \Big{|}_{\widetilde{A}} =  0 \,.
\eeq

\section{Bulk-to-boundary propagators}
\label{app:propagators}

We can be more explicit about the various propagators. Recall from \S\ref{sec3.2} the Euclidean bulk-to-boundary propagator written in $S^1\times \mathbb{H}^{d-1}$ coordinates:
\beq
K_{E}(r_B,\tau_B, Y_B| \tau, Y) = \frac{D_{\Delta}}{(-2 P\cdot X_B)^{\Delta}}=\frac{D_{\Delta}}{\left(-2 r_BY_B\cdot Y-2\sqrt{r_B^2-1} \cos(\tau_B -\tau)\right)^{\Delta}}
\eeq
while the Wightman propagators are given by
\beq
K_{\pm}(r_B,s_B, Y_B| s, Y) = \lim_{\epsilon \to 0^+}\frac{D_{\Delta}}{\left(-2 r_BY_B\cdot Y-2\sqrt{r_B^2-1} \cosh(s_B - s \mp i \epsilon)\right)^{\Delta}}
\eeq
\begin{figure}[h!]
\centering
\includegraphics[height=6cm]{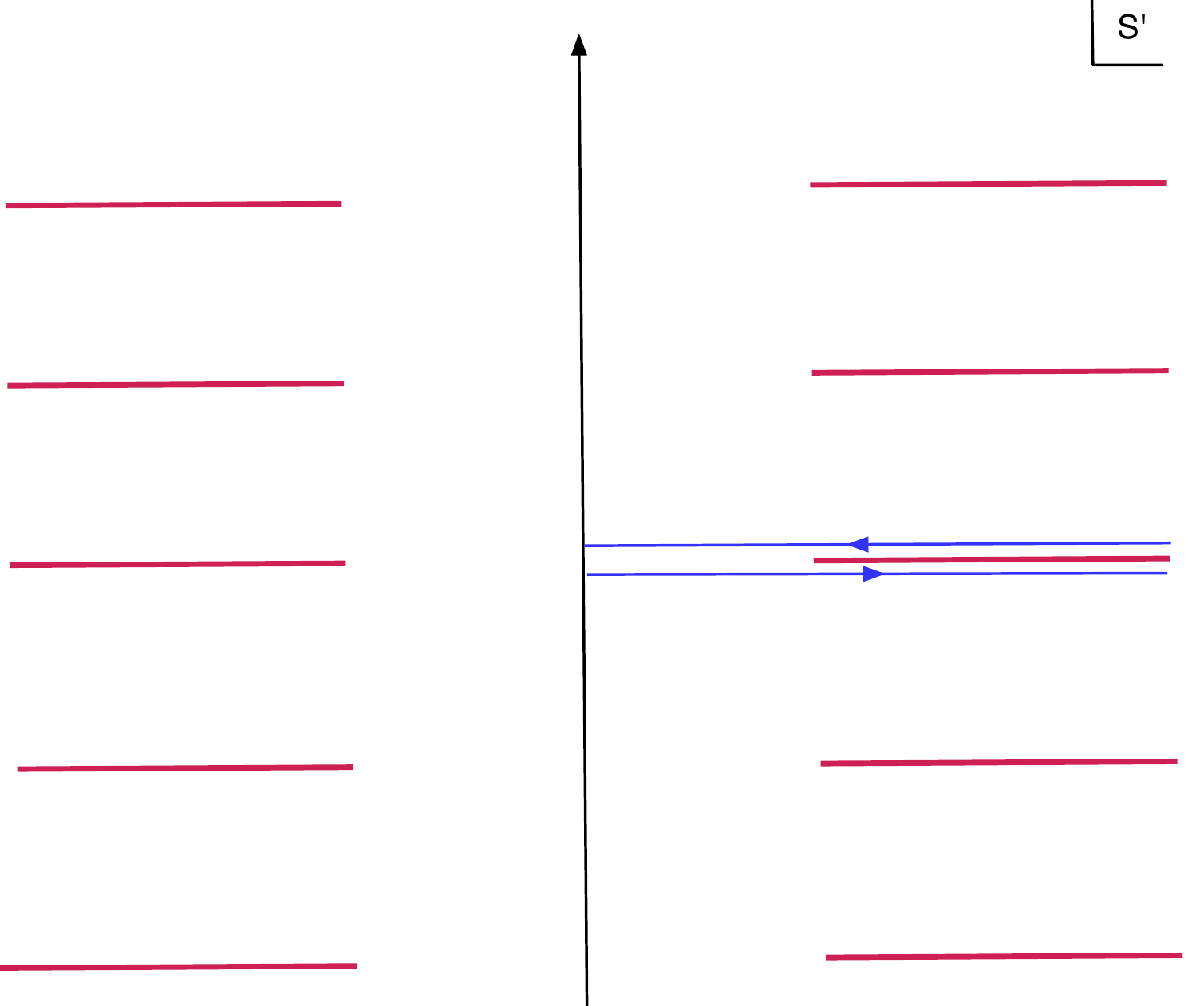}
\caption{\label{fig:sprime}\small{\textsf{The analytic structure in the complex $s'=s_B-s$ plane. The red lines are branch-buts. The blue line is the contour of integration.}}}
\end{figure}

We now briefly review why the retarded propagator approaches a delta function asymptotically and thus explain Eqs.\ \eqref{ae2}. It is clear that as $r_B \to \infty$, $K_R\to 0$ everywhere except when $(s_B,Y_B) \to (s, Y)$.  In this limit, the retarded propagator is the difference of two divergent quantities, which naively look like they cancel. But in order to see the delta function, we integrate the retarded propagator \eqref{eq:Kret} against a constant function
\small{
\beq \label{eq:IntDem}
i\int_0^{\infty} ds' \left(\frac{D_{\Delta}}{\left(-2 r_B Y_B\cdot Y-2 \sqrt{r_B^2-1}\cosh(s'-i\epsilon)\right)^{\Delta}}- \frac{D_{\Delta}}{\left(-2 r_BY_B\cdot Y-2\sqrt{r_B^2-1} \cosh(s'+i\epsilon)\right)^{\Delta}}\right)
\eeq
}\normalsize We can interpret this difference as a single integral along the contour that is shown in figure \ref{fig:sprime} together with the analytic structure in the complex $s'=s_B-s$ plane. By using the KMS condition, we can move the contour below the branch cut at $\mathrm{Im}(s')= -\epsilon$ to $\mathrm{Im}(s') = (2\pi -\epsilon)$. Then, using Cauchy's theorem we replace the integral along these two contours with the integral along a new contour running along the imaginary axis from $s=i\epsilon$ to $s=i(2\pi -\epsilon)$. (The piece at $s=\infty$ is exponentially suppressed, so we ignore it.) Now we see that this turns the above integral into precisely the Euclidean integral, which in the limit $r_B \to \infty$ we know evaluates to one.

\section{A simple example}
\label{app:example}

In this appendix we provide an explicit example to provide some intuition for the abstract discussion. We consider a homogeneous perturbation of AdS$_3$, which has been studied before in \cite{Lashkari:2015hha}. We merely reproduce their result here, but using the formalism and conventions developed in the main text.

\subsubsection*{Homogeneous perturbation in AdS$_3$}

Consider AdS$_3$ in Poincar\'e coordinates with the following perturbation:
\begin{equation}
  g_{\mu\nu}^{(0)} dx^\mu dx^\nu = \frac{1}{z^2} \left( -dt^2 + dx^2 + dz^2   \right) \,, \qquad h_{\mu\nu} dx^\mu dx^\nu  = \pert \left( dt^2 + dx^2 \right) \,,
\end{equation}
where $\pert$ is a perturbative parameter. The spherical boundary region of interest will be the interval $x\in [-1,1]$ at $t=0$.
This perturbation was previously considered in \cite{Lashkari:2015hha} in a similar context. Of course, in practical terms the computation is very simple in Poincar\'e coordinates. The point of the present section is to give a cross-check of our formalism.

We begin by recalling the gravitational result from \cite{Lashkari:2015hha}, i.e., the canonical energy \eqref{eqn:grav_result2}:
\begin{equation}\label{eq:delta2Sres}
\begin{split}
  \delta^{(2)}(E^{grav}_A - S_A^{grav}) + \int_{\Sigma_A}2 \xi_A^a E_{ab}^{(2)} \bs\epsilon^b&= \int_{\Sigma_A} {\bs \omega}_{grav} ( h, {\cal L}_{\xi_A}  h) + \int_{\widetilde{A}}  {\bs \chi}  \big(h , [\xi_A,V] \big)   \\
  &= -\frac{2 \pert^2}{15} + \frac{7 \pert^2}{45}
 = \frac{\pert^2}{45} \,.
\end{split}
\end{equation}

In the following, we will reproduce this result in our formalism. The transformation to our hyperbolic black hole coordinates reads
\begin{equation}
\begin{split}
  \ell_+ = \frac{(1+t)^2-x^2-z^2}{2z} \,, \qquad
  \ell_- = \frac{(1-t)^2-x^2-z^2}{2z} \,, \qquad
  u = \left( \frac{t^2 - (1-x)^2 - z^2}{t^2 - (1+x)^2 - z^2} \right)^{1/2} \,.
\end{split}
\end{equation}
The perturbation itself looks somewhat complicated in these coordinates. We only need its near-horizon expansion, which reads as
\small{
\begin{equation}
\begin{split}
&  h_{\mu\nu}  =  \frac{2\pert\,u^2}{(1+u^2)^4}
\left(
\begin{array}{ccc}
 \left(1+u^4\right) & -2 u^2 & 2 \left(1-u^2\right) \\
 -2 u^2 &  \left(1+u^4\right) & 2  \left(1-u^2\right) \\
 2 \left(1-u^2\right) & 2 \left(1-u^2\right) & 8  \\
\end{array}
\right) \\
&\quad+ \frac{2\pert\,u(1-u^2)}{(1+u^2)^5}
\left(
\begin{array}{ccc}
 \frac{2 u^2 \left(\ell_- \left(u^2-1\right)^2+2 \ell_+ \left(u^4+1\right)\right)}{u^2-1} & \frac{(\ell_-+\ell_+) u^2 \left(1-6 u^2+u^4\right)}{u^2-1} & \ell_+ (1-6u^2+u^4)- 8 \ell_- u^2 \\
 \frac{(\ell_-+\ell_+) u^2 \left(1-6 u^2+u^4\right)}{u^2-1} & \frac{2 u^2 \left(\ell_+ \left(u^2-1\right)^2+2 \ell_- \left(u^4+1\right)\right)}{u^2-1} &\ell_- (1-6u^2+u^4)- 8 \ell_+ u^2\\
\ell_+ (1-6u^2+u^4)- 8 \ell_- u^2 & \ell_- (1-6u^2+u^4)- 8 \ell_+ u^2 & \frac{4 (\ell_-+\ell_+) \left(1-6 u^2+u^4\right)}{1-u^2} \\
\end{array}
\right)\\
&\quad + \mathcal{O}(\ell_+,\ell_-)^2
\end{split}
\end{equation}
}
\normalsize

This perturbation is not in Hollands-Wald gauge. Instead, we find the following non-vanishing expressions at the surface $\widetilde{A}$:
\begin{equation}\label{eq:exHW}
\begin{split}
  \left( \nabla^{(0)}_\alpha  h^\alpha_{\;\bar{\a}} - \frac12 \nabla^{(0)}_{\bar{\a}}  h^\alpha_{\;\alpha} \right)_{\widetilde{A}} &=
   \frac{8 \pert  u^3 \left(u^2-1\right)^2}{\left(u^2+1\right)^5} \,,\\
 \cL_{\xi_A}  h |_{\widetilde{A}} &= \frac{8\pi\,  \pert  u^2 \left(u^4+1\right)}{\left(u^2+1\right)^4}
  \left(
\begin{array}{ccc}
 1 & 0 & \frac{1-u^2}{u^4+1} \\
 0 & -1 & \frac{u^2-1}{u^4+1} \\
 \frac{1-u^2}{u^4+1} & \frac{u^2-1}{u^4+1} & 0  \\
\end{array}
\right) \,.
\end{split}
\end{equation}
The vector $V^a$ which transforms $ h$ to Hollands-Wald gauge can be constructed explicitly:
\begin{equation}
 V =
\left( \begin{array}{ccc}
\frac{2 \pert  u \left(u^4+1\right)}{3 \left(u^2+1\right)^3} \\
\frac{2 \pert  u \left(u^4+1\right)}{3 \left(u^2+1\right)^3} \\
0
\end{array} \right)
+
\left( \begin{array}{ccc}
0\\
-\frac{2 \pert  u^2 \left(u^4+1\right)}{\left(u^2+1\right)^4}\\
\frac{\pert  u^2 \left(u^6+7 u^4-7 u^2-1\right)}{3 \left(u^2+1\right)^4}
\end{array} \right) \ell_+
+
\left( \begin{array}{ccc}
-\frac{2 \pert  u^2 \left(u^4+1\right)}{\left(u^2+1\right)^4}\\
0\\
\frac{\pert  u^2 \left(u^6+7 u^4-7 u^2-1\right)}{3 \left(u^2+1\right)^4}
\end{array} \right) \ell_-
+ \mathcal{O}(\ell_+,\ell_-)^2
\end{equation}
where we have already fixed various integration constants by demanding finiteness at $u=0$ and $u=\infty$. The Hollands-Wald expressions \eqref{eq:exHW} evaluated on the perturbation $ h + \cL_{V} g^{(0)}$ vanish. We now compute the relative entropy in different ways as a consistency check of our derivations.

\subsubsection*{Evaluation of CFT result}

We can now directly compute our CFT result \eqref{eq:d2Scft}, which reads\footnote{ We are assuming $\widetilde{C}_T = a^*$ here.}
\begin{equation}\label{eq:Int2}
\begin{split}
  \delta^{(2)} S (\rho || \rho_0) &= \int_{\Sigma_A} {\bs \omega}_{grav} ( h, {\cal L}_{\xi_A}  h) + \int_{{\cal H}^+} {\bs \omega}_{grav} ( h,- I(C_+) ) +  \int_{{\cal H}^-} {\bs \omega}_{grav} ( h, -I(C_-))   \,.
\end{split}
\end{equation}
The first integral over $\Sigma_A$ will, of course, match the result from above, $-\frac{2\pert^2}{15}$, since it is exactly the same term. Let us therefore investigate the integral over $\widetilde{A}$.
The integrals $I(C_+)$, $I(C_-)$ are given in this example by
\begin{equation}
\begin{split}
 I_{ab}(C_+) &= -2\pi
 \left( \begin{array}{ccc}
   0 & 0 & 0 \\
   0 & 2  ( h_{--} |_{\tilde A} ) + \ell_+ (\partial_+ h_{--} |_{\tilde A} ) &  ( h_{-u} |_{\tilde A} ) \\
   0 &  ( h_{u-} |_{\tilde A} ) & \ell_- (\partial_- h_{uu} |_{\tilde A} )
 \end{array} \right) +\ldots \,,\\
  I_{ab}(C_-) &= 2\pi
 \left( \begin{array}{ccc}
   2  ( h_{++} |_{\tilde A} ) + \ell_- (\partial_- h_{++} |_{\tilde A} ) & 0 &  ( h_{+u} |_{\tilde A} ) \\
   0 & 0 & 0 \\
   ( h_{+u} |_{\tilde A} ) & 0  & \ell_+ (\partial_+ h_{uu} |_{\tilde A} )
 \end{array} \right) +\ldots\,,
\end{split}
\end{equation}
where ``$\ldots$'' denotes higher order terms which don't contribute to the integral \eqref{eq:Int2}.
This gives the following symplectic fluxes on ${\cal H}^+$ and ${\cal H}^-$, respectively:
\small{
\begin{equation}
\begin{split}
  {\bs \omega}_{grav} ( h, -I^{(+)} ) \big{|}_{\ell_-=0} =
  \frac{4 \pert^2 u^4 \left(1+u^{10}+u^4(7 u^2+4)(u^2+1) -\ell_+ \,u\left(1+u^8-6 u^2 (u^4-u^2+1)\right)\right)}{ \left(u^2+1\right)^5 \left(\ell_+ u+u^2+1\right)^5} \, d\ell_+ du \\
   {\bs \omega}_{grav} ( h,- I^{(-)} ) \big{|}_{\ell_+=0} =
  \frac{4 \pert^2 u^4 \left(1+u^{10}+u^4(7 u^2+4)(u^2+1) -\ell_- \,u\left(1+u^8-6 u^2 (u^4-u^2+1)\right)\right)}{ \left(u^2+1\right)^5 \left(\ell_- u+u^2+1\right)^5}\, d\ell_- du
\end{split}
\end{equation}
}
\normalsize
Integrating these as in \eqref{eq:Int2} yields the same result as before, viz., \eqref{eq:delta2Sres}.

Finally, note that we can also compute the relative entropy in a third way by evaluating the final result of the CFT analysis, \eqref{cftresult}:
\begin{equation}\label{eq:Int1}
\begin{split}
  \delta^{(2)} S (\rho || \rho_0) &= \int_{\Sigma_A}\bs{\omega}_{grav} \left( h, \mathcal{L}_{\xi_A}  h\right) +\int_{\widetilde{A}}\Big( \bs{\chi} \left( h, [\xi_A,\cW_{(+)}] \right) + \bs{\chi} \left( h, [\xi_A,\cW_{(-)}] \right)\Big)  \,.
\end{split}
\end{equation}
Again, the first term is the same as previously. For the second term, we need to determine the vectors ${\cal V}_{(\pm)}$. It is clear that we need the latter at most to second order in the expansion around $\ell_+=\ell_-=0$. We find:
\begin{equation}
\begin{split}
 {\cal V}_{(+)} &= \left( \begin{array}{c} (V^+|_{\widetilde{A}}) + (\partial_- V^+|_{\widetilde{A}} ) \,\ell_- + \ldots \\ 0 \\ w_{(0)}(\ell_-,u) \end{array} \right) \,,\qquad
 {\cal V}_{(-)} = \left( \begin{array}{c} 0 \\ (V^-|_{\widetilde{A}}) + (\partial_+ V^-|_{\widetilde{A}} ) \,\ell_+ + \ldots  \\ w_{(0)}(\ell_+,u) \end{array} \right)\,.
 \end{split}
 \end{equation}
This yields
\begin{equation}
{\bs \chi} ( h, [\xi_A,{\cal V}_{(+)}]) \big{|}_{\ell_+ = \ell_- = 0} = {\bs \chi} (  h , [\xi_A,{\cal V}_{(-)}] )  \big{|}_{\ell_+ = \ell_- = 0}  = -\frac{2 \pert ^2 u^3 \left(3 u^4-2 u^2+3\right)}{3 \left(u^2+1\right)^6} \, du\,.
\end{equation}
Integration of this form again results in the same numerical value as in \eqref{eq:delta2Sres}.

\bibliographystyle{JHEP}
\bibliography{2OEE}
\end{document}